\documentclass[journal]{IEEEtran}
\usepackage{comment}
\usepackage{xcolor}
\usepackage{times,graphics,mathptm,epsfig,amsmath,xspace,endnotes,pifont,multirow,rotating,listings,amssymb,algorithmic,algorithm,color,soul,caption,nicefrac,adjustbox,todonotes,tabularx,mathtools,cite}
\usepackage{todonotes}
\usepackage{amssymb,amsfonts}
\usepackage{graphicx}
\usepackage{textcomp}
\usepackage{xcolor,url}
\usepackage{subcaption}
\def\BibTeX{{\rm B\kern-.05em{\sc i\kern-.025em b}\kern-.08em
    T\kern-.1667em\lower.7ex\hbox{E}\kern-.125emX}}

\usepackage{amsmath}
\usepackage{amsfonts}
\usepackage{amsthm}

\newtheorem{assumption}{Assumption}
\newtheorem{problem}{Problem}
\newtheorem{corollary}{Corollary}
\newtheorem{lemma}{Lemma}
\DeclareMathOperator*{\argmin}{argmin} 

  \usepackage{enumitem}   
\usepackage{diagbox}
\ifCLASSINFOpdf
\else
\fi

\usepackage{url}

\begin{document}

\title{FoReCo: a forecast-based recovery mechanism for real-time remote control of robotic manipulators}

\author{\IEEEauthorblockN{
Milan Groshev\IEEEauthorrefmark{1},
Jorge Mart\'{i}n-P\'{e}rez\IEEEauthorrefmark{1},
Carlos Guimar\~{a}es\IEEEauthorrefmark{2},
Antonio de la Oliva\IEEEauthorrefmark{1} and
Carlos J. Bernardos\IEEEauthorrefmark{1}}
\IEEEauthorblockA{
 \IEEEauthorrefmark{1}Universidad Carlos III de Madrid, Spain \\
 \IEEEauthorrefmark{2}ZettaScale Technology SARL, France
}}

\maketitle


\begin{abstract}
Wireless communications represent a game changer for future manufacturing plants, enabling flexible production chains as machinery and other components are not restricted to a location by the rigid wired connections on the factory floor. However, the presence of electromagnetic interference in the wireless spectrum may result in packet loss and delay, making it a challenging environment to meet the extreme reliability requirements of industrial applications. In such conditions, achieving real-time remote control, either from the Edge or Cloud, becomes complex.
In this paper,  we investigate a forecast-based recovery mechanism for real-time remote control of robotic manipulators (FoReCo) that uses Machine Learning (ML) to infer lost commands caused by interference in the wireless channel. FoReCo is evaluated through both simulation and experimentation in interference prone IEEE 802.11 wireless links, and using a commercial research robot that performs pick-and-place tasks. Results show that in case of interference,  FoReCo trajectory error is decreased by
x18 and \mbox{x2} times in simulation and experimentation, and that FoReCo is sufficiently lightweight to be deployed in the hardware of already used in existing solutions.
\end{abstract}

\begin{IEEEkeywords}
Robotic Manipulator, Wireless Remote Control, Machine Learning, IEEE 802.11, Interference.
\end{IEEEkeywords}

\section{Introduction}
\label{sec:introduction}
Real-time remote control and coordination of robot manipulators over a wireless network is seen as key enabler for future industrial applications~\cite{ind-teleop}, where a high level of flexibility, accuracy, data sharing and cost reduction are desired. While wireless is a must for mobile robots like Autonomous Guided Vehicles (AGVs), the implementation of wireless connections for robots manipulators also has many advantages such as greater flexibility, reduction of installation and maintenance costs, ease of scale, and less personnel exposure to hazardous situations~\cite{wcs}. Industry~4.0 scenarios will decide whether to use wireless technologies in the licensed spectrum, such as 5G New Radio~\cite{5g-nr};
or in the unlicensed spectrum, such as IEEE 802.11~\cite{IEEE80211}. 

Nowadays, industrial verticals implement IEEE 802.11 technologies for factory automation through commercial solutions such as Industrial WLAN (IWLAN) developed by Siemens. The low cost, good performance (e.g., low latency, high throughput), and extensive implementation in commercial equipment makes IEEE 802.11 a suitable candidate to fulfill the tight timing constraints of industrial automation. However, achieving the reliability, transparency and stability for real-time remote control required in many applications remains a critical challenge in IEEE 802.11. due to the highly unpredictable, unreliable, and interference prone wireless channel, which introduces delays, packet loss, jitter, throughput bottlenecks, and even loss of connectivity~\cite{802.11.problems}. The presence of packet collisions and electromagnetic (EM) interference in the shared medium results in delayed or even lost control commands. While the delayed delivery of control commands to the robot breaks the transparency of the remote control system (e.g., lag between the executed
remote control commands and the robot movements), the lost commands directly influence the stability resulting in deviation from the desired trajectory.

\begin{figure}[t]
\centerline{\includegraphics[width=\columnwidth]{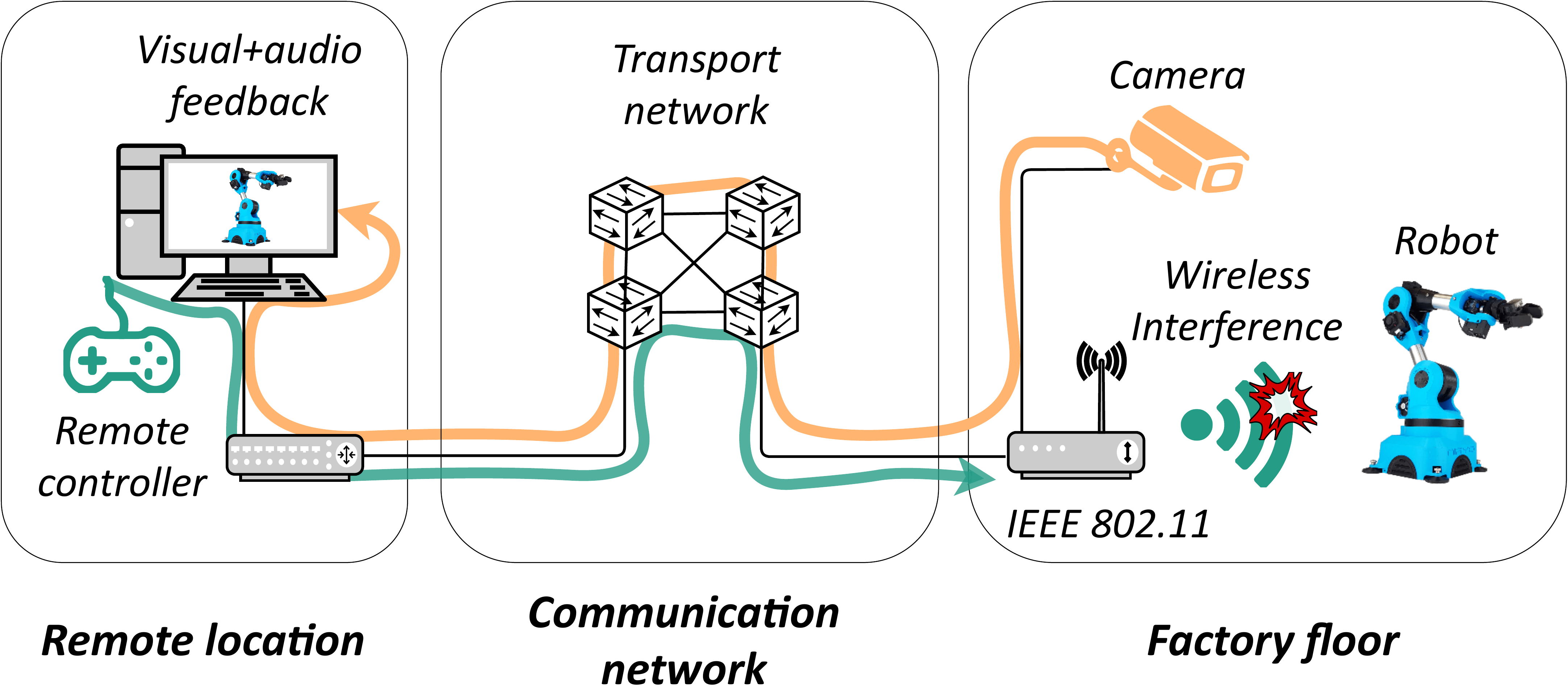}}
\caption{Diagram of a industrial robotic remote control.}
\label{fig:teleoperation-system}
\end{figure}
Mitigating the effect of delayed or lost commands in real-time remote
control robots is not new, as it has been tackled through the lens
of control theory in the robotics field~\cite{control-survey}.
However, most of these works do not consider the presence of EM interference, and assume that commands' delays are constant,
following sums of normal distributions, or following
first order Markov Processes. These assumptions do not consider the specifics of IEEE 802.11 Carrier-Sense Multiple Access with Collision Avoidance~(CSMA/CA) based Medium Access Control, 
as they do not
capture how the underlying back-off, and re-transmissions impact
the packet latency, nor the interference. Rather than making
assumptions about the commands' delay and designing a 
delay-tolerant control loop, this paper shifts the
focus and proposes a predictive control loop that infers the delayed or lost commands, and feeds them to the robot control loop. 

In this paper we propose FoReCo: a forecast-based recovery mechanism for real-time remote control of robotic manipulators. FoReCo is suitable for autonomous or human-assisted remote control of robot manipulators that perform repetitive tasks such as welding, materials handling, picking and packing, or assembly.
In case the robot does not receive a remote control command on time due to IEEE 802.11 collisions or EM interference, FoReCo ($i$) infers the delayed command; and ($ii$) injects it in the robot driver loop
so the operator does not perceive a misbehaviour in the remote control process.

This paper contributes to the state-of-the-art as follows:
\begin{itemize}
    \item We formulate an optimization problem to minimize the trajectory error due to delayed and lost real-time remote control commands;
    \item we propose FoReCo to infer delayed and lost commands using ML algorithms;
    \item we validate FoReCo via simulation using an IEEE 802.11 analytical model~\cite{80211analytical} that accounts for an interference source, and packet collisions due to channel neighbors; and
    \item we show experimentally that FoReCo works in a commercial research robotic arm, and mitigates the effects of electromagnetic interference created with a real jammer.
\end{itemize}

In the remainder of this paper, we
review the related work in~\S\ref{sec:related-work},
and formulate the problem statement in \S\ref{sec:problem}.
Then, in \S\ref{sec:solution} we present how FoReCo infers delayed/lost commands through ML.
In \S\ref{sec:wifi-model} we explain the analytical model
of IEEE~802.11 that we use to test FoReCo in simulated
scenarios with wireless interference.
Later in \S\ref{sec:results}, FoReCo is validated via simulation and real experiments. Finally, \S\ref{sec:discussion}~discusses the main insights from the obtained results, followed by conclusions and future directions in \S\ref{sec:conclusions}.

\section{Related work}
\label{sec:related-work}

There is a rising interest in the networking community
for providing support to Industry 4.0 cases in commercial
deployments. The goal is always to meet the reliability
that industrial processes require, as it is in the case
of remotely controlled robotic manipulators.
Under the umbrella of the 5G-Public-Private-Partnership
(5G-PPP) and
the European Research Council (ERC),
the European networking community has been
focusing on how 5G
and beyond 5G architectures can support the Industry 4.0.
5G-DIVE~\cite{5gdive} and 5Growth~\cite{5growth},
are two examples of platforms that manage the
adequate resource provisioning and allocation of services 
like remote control in Industry 4.0.
More recent European research
projects~\cite{5gppp-annual} as Daemon~\cite{daemon} and
Hexa-X~\cite{hexa-x} also provide support to
Industry~4.0 applications by bringing intelligence to
the network in order to meet strict constraints as reliability.

In the case of teleoperation applications
(i.e., applications providing remote control of systems),
there is a plethora of work in the robotic and mechatronic
literature on how to operate robotic manipulators.
Reference~\cite{teleoperationValencia}~presents a prototype
designed for the remote operation
of an industrial robot manipulator using augmented
reality, and~\cite{collission-avoid}
studies how to overcome, with the help of predictions,
the collision of a robotic manipulator with objects due to
remote operator errors.
Works as~\cite{imitation-learning}
and~\cite{natureteleoperation} propose teleoperated
robotic manipulators that assess complex and high
precision tasks with the help of ML assisted solutions,
and gravity compensation approaches, respectively.
However, the robotic and mechatronic literature
many times fails to consider the latency induced by network in teleoperated/remotely-controlled systems.
Indeed, none of the aforementioned works account for
the network latency in the problem formulation, nor in
the experimental stage, as authors control the robotic
manipulator with a computer directly attached to the robot.

Works as \cite{teleoperationValencia,collission-avoid,imitation-learning,natureteleoperation} introduce errors in remotely-controlled robotic manipulators when applied in networks that suffer from
high latencies or packet losses.
Also, they cannot rely on network platforms 
like~\cite{5gdive}, \cite{9422344}, \cite{daemon} and \cite{hexa-x} to overcome
such problems, as these platforms make a best-effort approach
by means of network resource allocation and life-cycle management. That is,
platforms as 5Growth~\cite{9422344} make the best to
allocate network
and computing resources for applications 
as~\cite{natureteleoperation}, but they do not assist
the remote-control application to recover when
control commands are delayed or lost in the network.

Therefore, it is up to the remote-control application
to decide how to react when control commands are delayed or lost. 
Mainly the robotic/mechatronic literature rely on
control theory to overcome issues produced by delayed
control commands.
Works as~\cite{TDPC-2,TDPC-3,TDPC-5,TDPC-6}
propose time domain passivity-based approaches,
in particular~\cite{TDPC-2} proposes to use
a two-layer and a switching
passivity-based~\cite{nonlinear-systems}
approach to deal with delays in the network.
Other solutions~\cite{wvbc-2,wvbc-3,wvbc-4}
cope with delay in remote-control using wave variable
passivity-based approaches~\cite{wvbc-1}.
Another option is to resort to adaptive and robust control
mechanisms to ensure the remotely-controlled robot
stability upon delayed commands,
as done
in~\cite{adaptive-1,adaptive-2,robust-adaptive,rbf}.
For example, \cite{rbf}~uses a Radial Basis Function Neural Network~\cite{rbf-nn-ref} based on Proportional Differential (PD) 
control~\cite{control-theory-book} to mitigate the effect
of external uncertainties and delayed commands in the
robotic manipulator.
However, the cited control-theory solutions in
robotic/mechatronic literature take unrealistic
assumptions about the remote control commands' delays.
In particular, \cite{TDPC-5} and \cite{adaptive-2}~assume
that the delay in the network is constant;
\cite{TDPC-3}, \cite{TDPC-6}, \cite{rbf}, \cite{wvbc-2} and \cite{wvbc-3}~assume that
delays are constant and with small variations;
and~\cite{TDPC-2}, \cite{robust-adaptive}, \cite{wvbc-4}, and \cite{adaptive-1} take the causality assumption for the delay,
i.e., the network delay cannot increase faster than
time\footnote{Following our notation, the causality assumption is expressed as: \mbox{$|\Delta(c_{i+1})- \Delta(c_i)|\leq |g(c_{i+1})-g(c_i)|$}}. All of the aforementioned assumptions
on network delay are not suitable for IEEE 802.11 wireless networks.

In this paper, we aim to improve the application and communication reliability by solving the problem from the
networking perspective. We use command predictions in order to recover from loss or delay of control packets.
There are also works in the state of the art that follow a similar
approach, in particular,
\cite{avg-2}~presents a control communication protocol that takes into account the wireless Signal to Noise Ratio (SNR)
and uses a reinforcement learning~\cite{sutton} approach~\cite{petar1996} to find the optimal speed of an AGV; and
\cite{avg-1}~proposes an AGV path tracking application using a Kalman filter to provide delay estimations for successful operation.
However, \cite{avg-2}~assumes that the success of the
wireless transmission is captured by a first-order
Markov process~\cite{first-order-markov}, and
\cite{avg-1}~assumes that the command delay in an IEEE 802.11
network follows a Gamma distribution. Both assumptions
neglect the presence of EM interference in the wireless
channel, as well as the back-off and re-transmission mechanisms
of IEEE~802.11 wireless channels that we investigate
in this paper. Moreover, both~\cite{avg-2} and~\cite{avg-1}
are solutions to enhance the reliability of real-time remotely
controlled AGVs in wireless networks, rather than
robotic manipulators.

\bigskip

The potential advantages from wireless remote control of a robot manipulator are significant, but realizing such systems over an IEEE 802.11 network remains a challenging task. To the best of our knowledge,
the state of the art disregards the presence of EM interference
in \mbox{real-time} remote control, and makes assumptions
about the remote control commands' delay that are not applicable to IEEE~802.11 wireless networks. To fill these gaps, we propose FoReCo, an ML-based solution that aims to minimize the robot trajectory error by predicting the missing remote control commands without making assumptions about the commands' delay.

\section{Problem statement}
\label{sec:problem}
A remote control system in industrial environments generally consists of three main parts: \textit{(i)} remote location; \textit{(ii)} communication network; and \textit{(iii)} factory floor (see~Fig.~\ref{fig:teleoperation-system}).
The remote site resides away from the factory floor, where a remote controller (fully autonomous or human-assisted) sends control commands to the factory robot in an open-loop fashion following a given frequency. Control commands which are sent every $\Omega$~[ms] have to transverse the transport network and wireless link to reach the robot.
Additionally, the remote site contains visual and audio feedback that provides similar conditions as those at the factory floor, and it is usually interconnected with wired technology to ensure reliability, a low-latency, and high bandwidth.

This work considers an IEEE 802.11 wireless link, as its
low price makes it an appealing solution for Industry~4.0.
However, the unlicensed IEEE 802.11 spectrum leads
to packet collisions, backoff times, and re-transmissions
that introduce delay in the control commands. That is, since
the moment a control command $c_i$ is generated $g(c_i)$,
up until the moment it is delivered to the robot $a(c_i)$, the
transport network and wireless link introduce a delay
{$\Delta(c_i)=a(c_i)-g(c_i)$}. Note that the latter is the addition of the delay introduced by the transport
network $\Delta_T(c_i)$, and the delay introduced by
the wireless link $\Delta_W(c_i)$, i.e.:
{$\Delta(c_i)=\Delta_T(c_i) + \Delta_W(c_i)$}.
In this paper, we make the following assumption on
the transport network delay:
\begin{assumption}
    The delay introduced by the transport network
    $\Delta_T(c_i)$ is upper bounded by a constant $D$:
    \begin{equation}
        \Delta_T(c_i) \le D,\quad \forall i
    \end{equation}
    \label{assum}
\end{assumption}
Since each network entity (e.g., switch or router) in the transport network
has finite queue sizes, the transport
network can be modeled as a Jackson network~\cite{jackson}.
Thus, we choose $D$ as a constant
above than the summation of waiting times and processing
time at each queue within the remote control path.

\begin{figure}[!t]
\centerline{\includegraphics[width=\columnwidth]{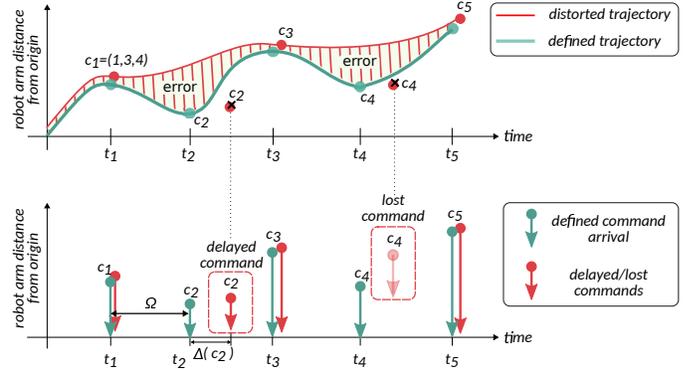}}
\caption{Impact of delayed and lost commands in the robot trajectory.}
\label{fig:problem-statement}
\end{figure}
However, even if $D$ is very small,
the delay introduced by the IEEE 802.11 link $\Delta_W(c_i)$
might lead to a laggy behavior in the remotely controlled robot. This means that the remote controller will experience a lag in between the time it moves the
controller, and the time the robot moves.
Since remotely controlled robots can only tolerate waiting
for $\tau$ milliseconds to receive the next command,
if $\Delta(c_i)>\tau$ the command $c_i$ will exceed the
tolerated delay, and the robot will not execute it.
Thus, it is necessary that the control commands
delays satisfy $\Delta(c_i)\le \tau$.

Control commands are sent every \mbox{$\Omega$ ms} and the robot expects to receive those commands in the same interval (\mbox{$\Omega$ ms}) for smooth operation. However, due to the network delay
the next control command $c_{i+1}$ might not arrive until
\mbox{$\Omega+\Delta(c_{i+1})$ ms} have passed. 

Overall, the robot will not execute commands
that arrive out of time $\Delta(c_i)>\tau$,
or are lost {$\Delta(c_i)\to\infty$}.
Upon any of these situations, the command is not executed, resulting in a deviation from the ideal trajectory that
the robot should follow (see Fig.~\ref{fig:problem-statement}).
Note that the remote controller will notice this error
in the real trajectory via the visual feedback that it receives
from the factory floor (see Fig.~\ref{fig:teleoperation-system}).
Thus, it is necessary to recover the discarded packets
to minimize the error in the real trajectory.

\noindent\begin{minipage}{\columnwidth}
\begin{problem}
Given the random variables $\Delta(c_i)$,
a distance
{$d: \mathbb{R}^d\times\mathbb{R}^d\mapsto\mathbb{R}$},
a tolerance $\tau$, and
a record of the last $R$ commands;
find {$f:\mathbb{R}^{d} \times \underbrace{\ldots}_{R-2} \times\mathbb{R}^{d} \mapsto \mathbb{R}^d$} to solve

\begin{align}
        &\min_{\hat{c}_N} &\quad &  \lim_{N\to\infty} \frac{1}{N} \sum_i^N d(\hat{c}_i,c_i)\hfill  \label{eq:min} \\
        &\text{s.t.} & & \hat{c}_i = f\left( \{\hat{c}_j\}_{i-R}^{i-1} \right)\mathbf{1}_{\Delta(c_i)>\tau} + c_i\left[1-\mathbf{1}_{\Delta(c_i)>\tau} \right],  \,\forall i \label{eq:command}
\end{align}
\label{problem}
\end{problem}
\end{minipage}
With $\mathbf{1}_{\Delta(c_i)>\tau}=1$ if command $c_i$
is delayed more than $\tau$ ms, and zero otherwise.
Problem~\ref{problem} targets to~find a function
$f$ to derive those commands that did not arrive on time.
Additionally, the derived commands $\hat{c}_i$ should
minimize the error (i.e., the distance $d(\hat{c}_i,c_i)$)
with respect to the~original command $c_i$ sent by the
remote controller. That is, the objective of~\eqref{eq:min}
is to minimize the dashed error region in
Fig.~\ref{fig:problem-statement}.

\section{Forecast-Assisted Remote Control (FoReCo)}
\label{sec:solution}

In this section, we present FoReCo as a forecast-based recovery mechanism to minimize
the trajectory error of remotely controlled robots via
wireless connectivity.

\subsection{The FoReCo Building Block}
\label{subsec:foreco-building-block}

As discussed in \S\ref{sec:problem}, whenever the command
delay exceeds the tolerance {$\Delta(c_i)>\tau$}, the robot
considers the command $c_i$ to be outdated and does
not execute it. Depending on the robot, the absence of the
command $c_i$ may result in the robot stopping, or keep
feeding the prior command $c_{i-1}$ to the robot control loop,
which is implemented
with solutions as Proportional-Integral-Derivative (PID)
controllers (see~\cite{moveit}). Either way, the command
$c_i$ will not be executed and the robot trajectory will
deviate from the expected, i.e., the trajectory executed by
the remote controller.
It is at this point that FoReCo predicts the command $c_i$ that has not arrived on time and transparently triggers its execution into the robot.
Hence, FoReCo stands as a complementary solution for any
remotely controlled robot using a wireless link, while being
agnostic to the implemented robot controller (control theory-based or not).

To predict control commands out of time,
FoReCo follows a machine learning (ML) based approach, which has been proven to be effective with intention prediction and estimation of future trajectories of objects, such as vehicles, bikes, and humans.
The learning model consist of predicting incoming control commands {$c_i, c_{i+1}, c_{i+2},\ldots$}
with the help of the prior {$c_{i-1},c_{i-2},\ldots$} commands.
To do so, we advocate for a ML
based methodology due to ($i$) the repetitive nature of
the industrial tasks performed by remotely operated robots; and
($ii$) the difficulty to solve this problem with traditional
dynamic programming algorithms. 

\begin{figure}[t]
    \centering
    \includegraphics[width=\columnwidth]{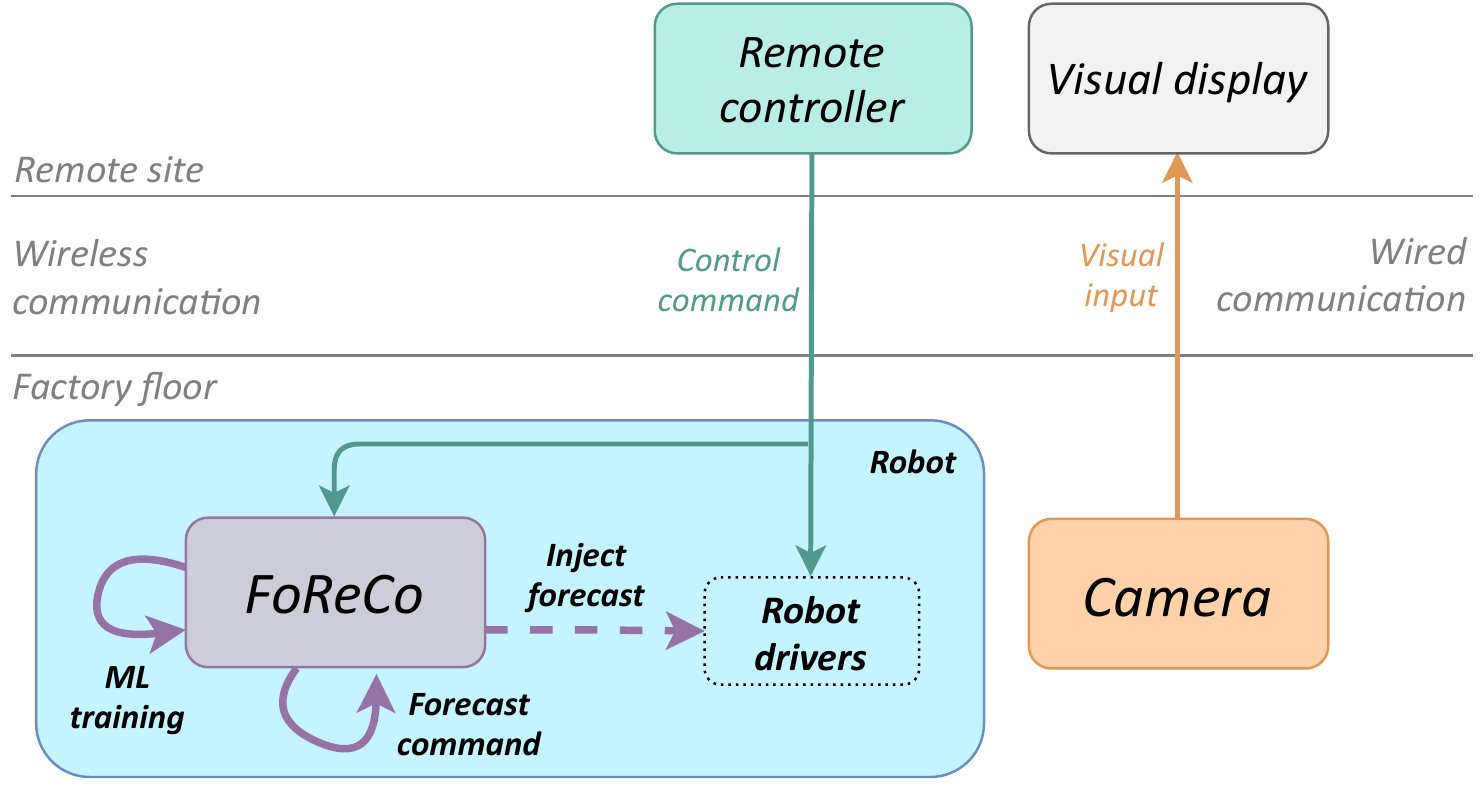}
    \caption{FoReCo building block and remote control system.}
    \label{fig:foreco-integration}
\end{figure}
Fig.~\ref{fig:foreco-integration} shows the conceptual components of the network control system we use to remotely control a
robot (in-line with Fig.~\ref{fig:teleoperation-system}).
The system shows the details of the interactions between the remote site and the factory floor over a communication channel. First, a real-time video stream of the robot is presented to a visual display over a wired communication channel. For simplicity, we assume that the uplink channel is error and delay-free and the video input is delivered to the remote operator immediately. The remote controller, with the help of the visual input, sends control commands over the wireless communication channel, and the commands are received by both the robot and FoReCo.
With the received commands, FoReCo performs two actions:
\begin{enumerate}
    \item \textbf{ML training}: to solve~Problem~\ref{problem},
    FoReCo resorts to ML to derive {$f\left(\{c_j\}, \Vec{w}\right)$},
    with $\Vec{w}$ being the weights to learn (see \S\ref{subsec:studied-algos}).
    To obtain $\Vec{w}$, FoReCo creates a dataset
    (see Fig.~\ref{fig:dataset})
    with the commands it receives
    from the remote controller. The dataset contains a
    history of $H$ commands, and FoReCo uses $\alpha H$
    of them for training, and $\beta H$ for testing; with
    {$\alpha+\beta=1$}. As in Problem~\ref{problem}, the
    training procedure aims to minimize the distance between
    predicted commands $\hat{c}_i$, and the ones sent by
    the remote operator $c_i$. Hence, FoReCo trains its
    ML solution $f(\{c_j\},\Vec{w})$ s.t.:
    \begin{equation}
        \min_{\Vec{w}} \frac{1}{\alpha H} \sum_i^{\alpha H} d\left( c_i, f\left(\{c_j\}_{i-R}^{i-1}, \Vec{w} \right) \right)
        \label{eq:foreco-train}
    \end{equation}
    With the obtained weights $\Vec{w}$, FoReCo tests
    the ML predictions accuracy in the testing set
    $\beta H$.
    
    \item \textbf{Command forecast and injection}: FoReCo
    awaits a control command $c_i$ each
    $\Omega$~ms, and it triggers the forecasting
    if the next command
    $c_{i+1}$ arrives latter than {$a(c_i)+\Omega+\tau$}.
    In this case, FoReCo will forecast the next command as
    {$\hat{c}_{i+1}=f\left(\{\hat{c}_j\}_{i-R}^i, \Vec{w}\right)$}, i.e., using the ML solution $f$ and
    the weights $\Vec{w}$ obtained from the training stage.
    The forecast command $\hat{c}_{i+1}$ is injected in
    the robot drivers (as illustrated in
    Fig.~\ref{fig:foreco-integration}) with the latter assuming that it received a command on time. In the
    case a command arrives on time
    {$a(c_{i+1})\leq a(c_i)+\Omega+\tau$}, FoReCo will just
    store the command in the dataset and later use it for
    training and forecasting purposes. Note that we
    refer to $\hat{c}_i=c_i$ if the command arrived
    on time $\Delta(c_i)\leq\tau$, so it satisfies
    the constraint stated in~\eqref{eq:command}.
    Thus, the forecasting receives as input
    {$\{\hat{c}_j\}_{j-R}^i$} commands that arrived on time,
    and the forecasts of those commands that did not arrive
    on time.
\end{enumerate}

\subsection{Studied ML Forecasting Algorithms}
\label{subsec:studied-algos}
In the following, the selected ML algorithms used to implement FoReCo are described, as well as how they perform
the command forecasting.

FoReCo is designed to forecast 
commands of remotely controlled robotic arms as the one
in Fig.~\ref{fig:teleoperation-system}. Each command
consists of $d$ joints (remember \mbox{$c_i\in\mathbb{R}^d$})
that move together to shift the arm manipulator position,
so the latter reaches the object of interest. Each command coordinate \mbox{$c_i=(c_i^1,\ldots,c_i^d)$} 
represents the rotation angle or shift of a joint, depending on
the joint nature. FoReCo considers the following ML algorithms:
\begin{itemize}
    \item \textbf{Vector Autoregression (VAR)}:
    this regression solution is designed to predict
    multi-dimensional time-series with correlation across
    dimensions. This is the case of robotic arms, whose
    joint coordinates typically present correlation
    \mbox{$\forall i,\exists k,m: c_i^k\sim c_i^{k+m}$};
    as they have to move together to reach and grab an object.
    VAR derives the prediction of a command as follows:
    \begin{equation}
        \hat{c}_{i+1}^k=f^k\left(\{\hat{c}_j\}_{i-R}^i, \Vec{w}\right) = b^k + \sum_{l=1}^d\sum_{j=i-R}^i w_{i,j}^l\cdot\hat{c}_j^l,\quad k\leq d
        \label{eq:var}
    \end{equation}
    with $b^k$ being the bias for the $k$\textsuperscript{th}
    coordinate, $w_{i,j}^l$ the regression weights,
    and $f^k(\cdot)$ denoting the
    $k$\textsuperscript{th} coordinate of the resulting
    prediction. Both $b^k,w_{i,j}^l$ elements lie within the
    weight vector $\Vec{w}$.
    
    \item \textbf{Sequence to sequence (seq2seq)}~\cite{seq2seq}:
    this ML solution is based on a Neural Network (NN)
    that receives as input a sequence and produces an output,
    that in our case is just a single output. These
    seq2seq models are known as many-to-one, as we feed
    it with a sequence of past commands $\{\hat{c}_j\}_{i-R}^i$
    to produce a single one $\hat{c}_i$. The seq2seq
    architecture we use has:
    ($i$) an encoder layer of 200 Long Short-Term Memory (LSTM) neurons with Rectifier Linear Unit (ReLU) activations;
    ($ii$) and a decoder layer of 30 LSTM neurons, also
    with ReLU activations.
    The motivation behind the use of a seq2seq solution
    is to learn and encode the characteristics of
    robot movements, so the decoder layer ``interprets''
    the encoded characteristics and guess the next
    command $\hat{c}_{i+1}$. Analytically, the seq2seq
    encoder and decoder layers are represented as follows:
    \begin{align}
        e_{i}^k = & \phi^k\left( W_0\hat{c}_i + W_1 a_{i-1} + W_2 m_i  \right), \quad k\leq d\label{eq:encoder}\\ \nonumber
        \hat{c}_{i+1}^k = & f^k\left(\{\hat{c}_j\}_{i-R}^i, \Vec{w}\right) =\\
        =& \phi^k\left( W_4 e_i^k + W_5 a_{i-1}' + W_6 m_i'  \right), \quad k\leq d
        \label{eq:decoder}
    \end{align}
    with $\phi(x)$ denoting the ReLU function\footnote{
    $\phi(x)=0, x\le0$ and $\phi(x)=x$ otherwise},
    $W_i$ denoting weight matrices (whose values
    are unrolled to derive the weight vector $\Vec{w}$),
    $a_{i-1}, a_{i-1}'$
    being the output of the LSTM activation units,
    and $m_i,m_i'$ referring to the memory cells of the encoder
    and decoder; respectively.
    
    \item \textbf{Moving Average (MA)}:
    we resort to this algorithm to have a benchmark for
    VAR and seq2seq. The MA derives the command prediction
    using:
    
    \begin{equation}
        \hat{c}_{i+1} = f^k\left(\{\hat{c}_j\}_{i-R}^i, \Vec{w}\right) = \frac{1}{R} \sum_{j=i-R}^i \hat{c}_j
        \label{eq:ma}
    \end{equation}
\end{itemize}

Note that FoReCo is flexible to support other forecasting algorithms, which can be integrated in a modular fashion. 

\subsection{Training of Selected ML Forecasting Algorithms}
\label{subsec:training}
Next, we detail how FoReCo trains the selected ML algorithms described above.

\begin{itemize}
    \item \textbf{VAR training}: to train the VAR algorithm,
    we resort to Ordinary Least
    Squares (OLS), i.e., the weights are computed as follows:
    \begin{equation}
        \Vec{w} = \argmin_{\Vec{w}} \sum_i^{\alpha H}\sum_k^d \left( c_i^k - f^k(\{c_j\}_{i-R}^{i-1}, \Vec{w}) \right)^2
        \label{eq:ols}
    \end{equation}
    over the training portion of the dataset $\alpha H$.
    Here $f^k(\cdot)$ refers to~\eqref{eq:var}.
    Note that minimizing the summation in~\eqref{eq:ols} is
    equivalent to minimizing the expression
    in~\eqref{eq:foreco-train}, taking as distance
    $d(c_i, \hat{c}_i)=\sum_k \left( c_i^k - \hat{c}_i^k \right)^2$.
    
    \item \textbf{seq2seq training}: we train the seq2seq
    solution using Adam, a
    stochastic optimization method invariant to
    small gradients, as it is the case of our experimental
    study, e.g., a the 3\textsuperscript{rd} robot joint
    takes values like $c_i^3=0.001$. We use Adam to iteratively
    minimize the error of the forecasts within a batch
    $B_i$ of commands from the training set, i.e.,
    $B_i<\alpha H$. Hence, at each training step Adam
    minimizes:
    \begin{equation}
        \min_{\Vec{w}_t}~
        l\left(\{c_j\}_{j=0}^{B_i},\Vec{w}_t\right) = \sum_i^{\alpha H}\sum_k^d \frac{\left( c_i^k - f^k(\{c_j\}_{i-R}^{i-1}, \Vec{w}_t) \right)^2}{B_i}
        \label{eq:adam}
    \end{equation}
    with $l(\cdot)$ denoting the loss function, and $\Vec{w}_t$
    the updated weight vector at the step $t$ of the training
    stage. Note that minimizing~\eqref{eq:adam} is equivalent
    to minimizing~\eqref{eq:foreco-train} over the batch $B_i$,
    rather than the whole training dataset $\alpha H$, taking
    the sum of squared distances. At each step the weights
    are updated as follows:
    \begin{align}
        \Vec{w}_{t+1} &= \Vec{w}_t - \eta\frac{m_{t+1}}{1-\beta_1^{\alpha H}} \frac{1}{\sqrt{\frac{v_i}{1-\beta_2^{\alpha H}}} + \varepsilon}\\
        m_{t+1} & = \beta_1 m_t + (1-\beta_1) \nabla_{\Vec{w}} l\left(\{c_j\}_{j=0}^{B_i},\Vec{w}_t\right)\\
        v_{t+1} & = \beta_2 v_t + (1-\beta_2) \left[ \nabla_{\Vec{w}}l\left(\{c_j\}_{j=0}^{B_i},\Vec{w}_t\right)  \right]^2
    \end{align}
    with $m_t,v_t$ being the estimates of the first and
    second moment of the loss function gradient $\nabla_{\Vec{w}}l(\cdot)$, $\eta$ the step size, and
    $\beta_1,\beta_2,\varepsilon$ other hyper-parameters.
    
\end{itemize}

\section{IEEE 802.11 with Electromagnetic  Interference}
\label{sec:wifi-model}
So far we have discussed how FoReCo
works in~\S\ref{subsec:foreco-building-block},
and the ML solutions that we consider to assess the
command forecasting in~\S\ref{subsec:studied-algos}
and~\S\ref{subsec:training}, respectively. In this section
we explain the analytical model we consider to derive the
delay that control commands experiment $\Delta_W(c_i)$ in
IEEE~802.11 wireless links under EM interference. The analytical
model is latter used in~\S\ref{subsec:simulation} to derive
the $\Delta_W(c_i)$, and assess the performance of
FoReCo in a simulated scenario as close as possible to
real IEEE~802.11-based real-time remote control.

In this paper we resort to the analytical 
model presented in~\cite{80211analytical} to derive wireless
delays. This work models the MAC layer of IEEE~802.11 with CSMA/CA,
and studies how neighboring nodes, and non-IEEE~802.11 interfering
sources impact the wireless delay. The work is based
on a refinement~\cite{pham} of Bianchi's characterization
of IEEE~802.11~\cite{bianchi}. The particularity is that
\cite{802.11.problems}~extends the underlying Markov chain
to also capture the presence of an interference source
that is active during $T_{if}$ transmission slots, and
emits with a probability $p_{if}$.
The proposed model also captures both
the back-off mechanisms and re-transmissions (RTX) of frames
upon collision in the IEEE~802.11 wireless link.

With the aforementioned model, \cite{80211analytical}~obtains the
steady-state vector of each state, in particular, they
derive the probability that a frame
has to be transmitted
after $j$ unsuccessful re-transmissions, which is denoted
as $a_j$. Moreover, \cite{80211analytical}~also derives
$\mathbb{E}_j\left[ \Delta_W(c_i) \right]$, that is,
average delay that the command $c_i$ experiences in the
wireless transmission after $j$ unsuccessful re-transmissions.
%
Based on such expression, we derive in the Appendix
some theoretical results around the analytical model given in~\cite{80211analytical}, that give some insights about the
delay of control commands.
In particular, the theoretical results in the Appendix
conclude that in the considered IEEE~802.11 scenario:
\begin{enumerate}[label=\textbf{\roman*)}]
    \item $\Delta(c_i)$ is only bounded on average, but
    not always (see Lemma~\ref{lemma:upper-bound});
    \item $\Delta(c_i)$ diverges (see
    Corollary~\ref{corol:unbound}); and
    \item the causality assumption does not apply
    (see Corollary~\ref{corol:no-causality}).
\end{enumerate}
Hence, the delay assumptions taken in the solutions
presented in~\S\ref{sec:related-work} do not hold.
In other words, we cannot bound the delays that the
remote control commands $c_i$ are experiencing.
Still, we resort to the analytical model presented
in~\cite{80211analytical}, as such unbounded delay behaviours are
actually realistic in IEEE~802.11 scenarios upon presence
of interference sources. 

\begin{figure}
    \centering
    \includegraphics[width=\columnwidth]{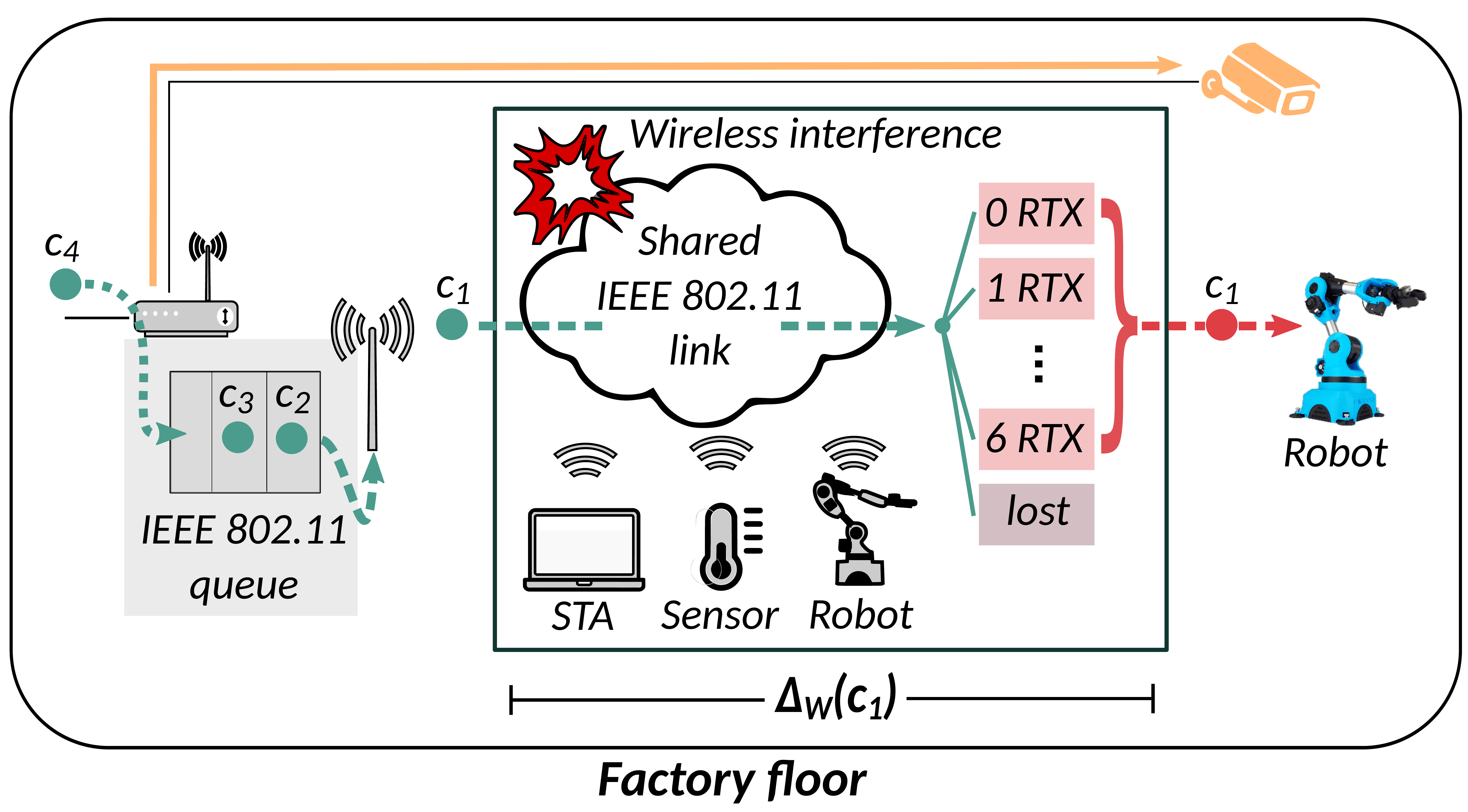}
    \caption{Impact of wireless interference, retransmissions (RTX), and factory devices in the delay $\Delta_W(c_i)$ that control commands experience in an IEEE~802.11 link.}
    \label{fig:hyperexp}
\end{figure}
To derive the value of $\Delta_W(c_i)$ we 
follow~\cite{80211analytical} and model the transmission
of control commands $c_i$ over IEEE~802.11 wireless links
as a queuing model. From the problem statement
formulation presented in~\S\ref{sec:problem}, we know that control
commands have an arrival rate $\tfrac{1}{\Omega}$. These
commands are queued in the IEEE~802.11 access point
before they are transmitted to the shared wireless link,
and they may require up to 6 re-transmissions before
they are received by the robotic arm
(see Fig.~\ref{fig:hyperexp}). Or even worse, the control
commands may be lost due because the RTX limit is exceeded,
due to collisions with packets sent by other devices
in the factory floor, or because of wireless interference
(see Fig.~\ref{fig:hyperexp}). Depending on the number
of RTX, the control command delay $\Delta_W(c_i)$ will
be higher or lower. This system behaves as an
$G/HEXP/1/Q$ queuing model, with $Q$ being the length of the
access point queue, and the service rates of the
hyperexponential distribution corresponding to the average delay
that control commands see after $j$ RTX, i.e., 
$\tfrac{1}{\mathbb{E}_j\left[ \Delta_W(c_i) \right]}$.

Given this $G/HEXP/1/Q$ queuing model, we can derive
$\Delta_W(c_i)$ in the desired IEEE~802.11 wireless scenario
accounting for the number of transmitting devices and
the probability and time that the wireless interference is
active. These are the delay values used in the
simulation scenarios in~\S\ref{subsec:simulation}, and
we derive them using the CIW discrete event simulation
library~\cite{ciw}.

\section{Results}
\label{sec:results}
In order to evaluate FoReCo, we consider a realistic industrial application where a robot manipulator is remotely controlled to perform a pick and place task. This remote control application allows us to select the most suitable forecasting algorithm for FoReCo (see~\S\ref{subsec:studied-algos}), and to evaluate a prototype implementation of FoReCo under simulation (see~\S\ref{subsec:simulation}) and experimental scenarios (see~\S\ref{subsec:experimental}). It is worth mentioning that in this section we compare the performance of FoReCo with the baseline Niryo controller mainly because of the lack of state-of-the-art remote controllers for robot manipulators. We leave the comparison to more recent AVG approaches, e.g., learning-based AVG controllers~\cite{avg-2}, for follow-up work, where the AVG controllers need to be adapted for robot manipulators in order to employ their benefits. 

\subsection{Testbed setup and dataset collection}
\label{subsec:setup}
\begin{figure}[t]
    \centering
    \includegraphics[width=\columnwidth]{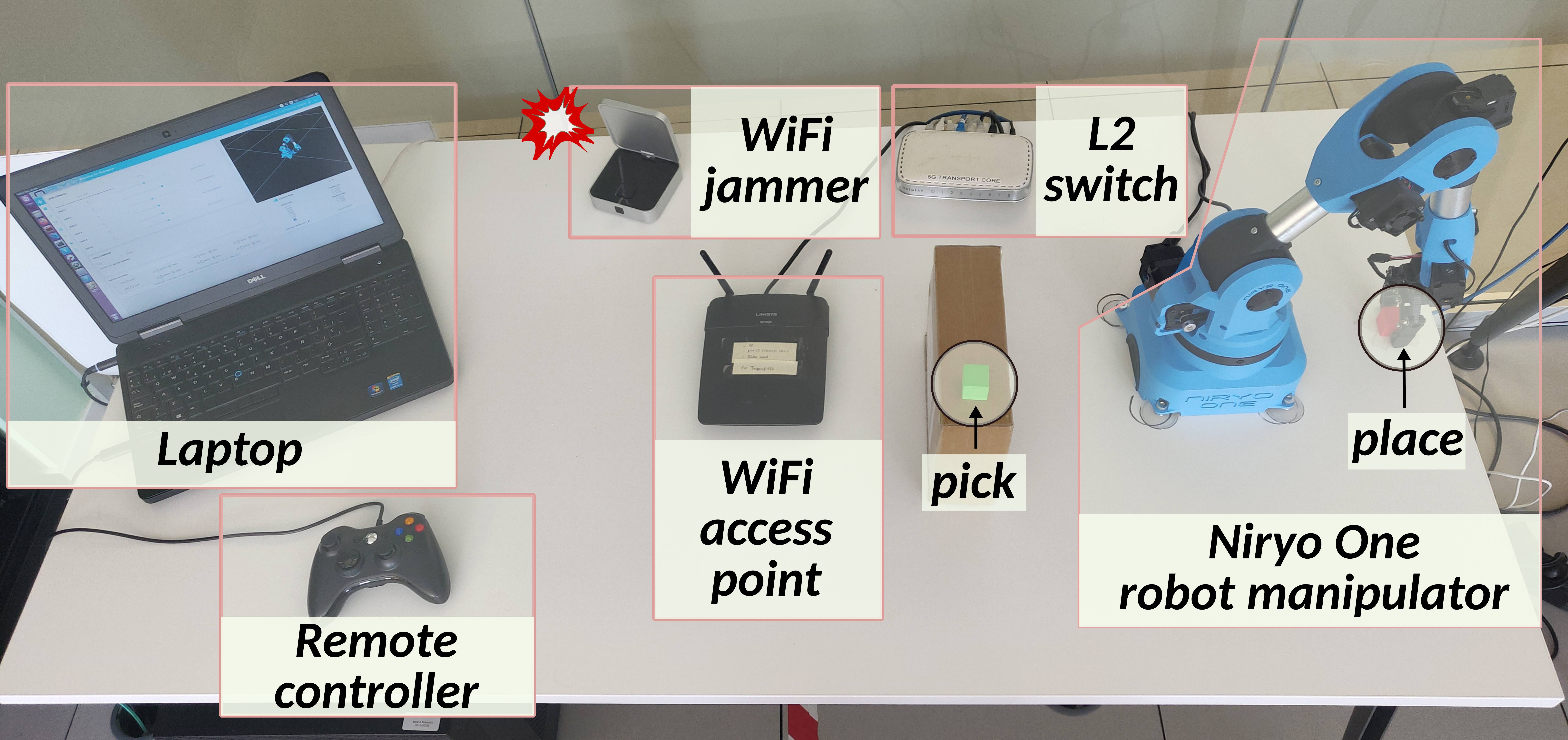}
    \caption{Testbed setup.}
    \label{fig:experimental-scenario}
    \vspace{-1em}
\end{figure}

Fig.~\ref{fig:experimental-scenario} shows the experimental testbed, built in the 5TONIC laboratory\footnote{https://www.5tonic.org} that is composed of: a 6-axis Niryo One robotic manipulator, a 2.4 GHz IEEE 802.11 access point (AP), an L2 switch, a 2.4 GHz Silvercrest Wireless Transmitter that is used as WiFi jammer, and a joystick that is connected to a laptop with 8GB RAM and 4CPU@2.4GHz via USB. The communication between the robot and the joystick is comprised of a wireless link from the robot to the AP and an Ethernet link from the AP to the laptop using the L2 switch.
The Niryo One robotic arm is equipped with a \mbox{Raspberry~Pi 3} with a 1.2 GHz 64-bit CPU, 1GB RAM, and a IEEE 802.11n interface. The robot maximum speed is 0.4 m/s for the steeper axes and 90°/s for the servo axis with a configured Robot Operating System (ROS) control command frequency of 50Hz (i.e., \mbox{$\Omega=20$ ms}) and command moving offset of 0.04 rad. The remote control and robot system are based on ROS version 1.
The Niryo One ROS stack expects to receive control commands
each $\Omega$ ms, and considers that a packet did not arrive
on time otherwise, i.e., the tolerance is $\tau=0$.
Niryo One ROS stack uses the prior command $\hat{c}_{i+1}=c_i$
in case $\Delta(c_{i+1})>\Omega$. Every received command is
passed to the Niryo motion planing layer (MoveIt),
which uses Proportional–Integral–Derivative (PID) control.

\begin{figure}[t]
    \centering
    \includegraphics[width=\columnwidth]{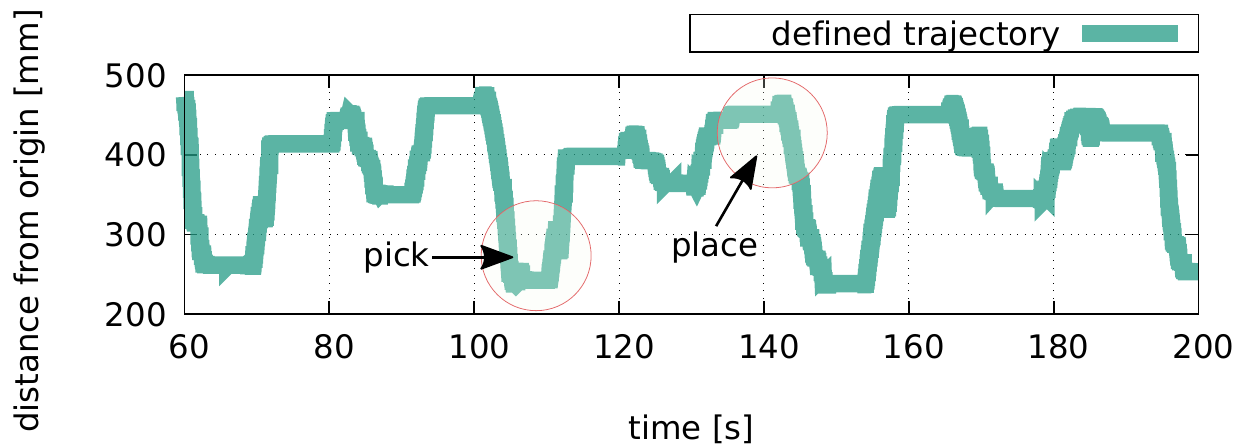}
    \caption{Robot trajectory dataset with pick and actions of an inexperienced operator.}
    \label{fig:dataset}
\end{figure}

Fig.~\ref{fig:dataset} shows part of the dataset created by performing pick and place actions. The pick and place actions were manually repeated 100 times by two different human operators, an experienced and inexperienced human operator resulting in the creation of two separate datasets. To do so, they used the joystick as a remote controller, issuing a new control command every \mbox{20 ms}.
The inexperienced/experienced operators datasets'
contain \mbox{$H=187109$} commands.
Both datasets store the joint states $c_i$ of the robot manipulator
under ideal network conditions, i.e., low latencies and
absence of packet collision. To achieve such conditions,
the datasets were obtained using
Ethernet to send the remote controller commands. The experienced dataset was used to train the ML models while the inexperienced data was used for remote control and testing. In this way we ensure that the trained ML model operates on data that is tightly related but not exactly the same as the training data. 


\begin{figure}[b]
    \centering
    \includegraphics[width=\columnwidth]{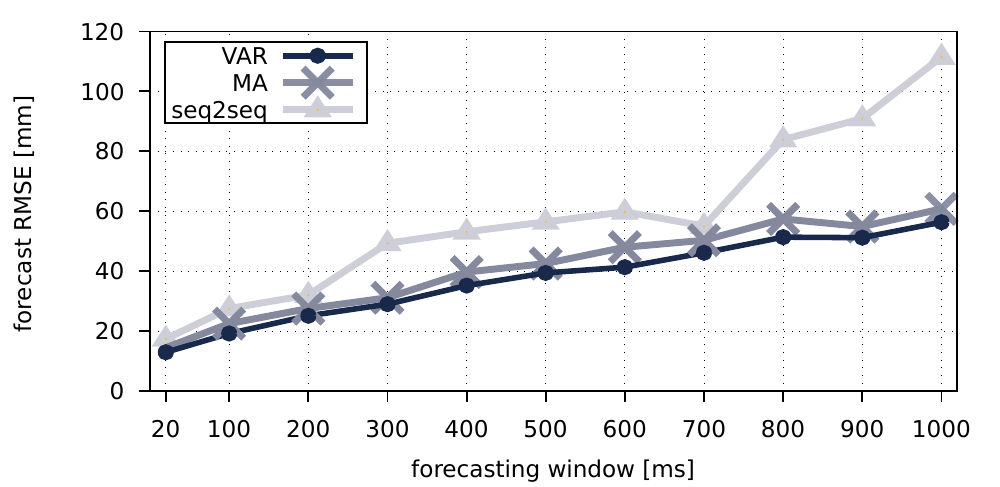}
    \caption{Forecast accuracy for different forecasting windows.}
    \label{fig:impact-ahead}
\end{figure}

\begin{figure*}[th]
     \centering
     \begin{subfigure}[b]{0.3\textwidth}
         \centering
         \includegraphics[width=\textwidth]{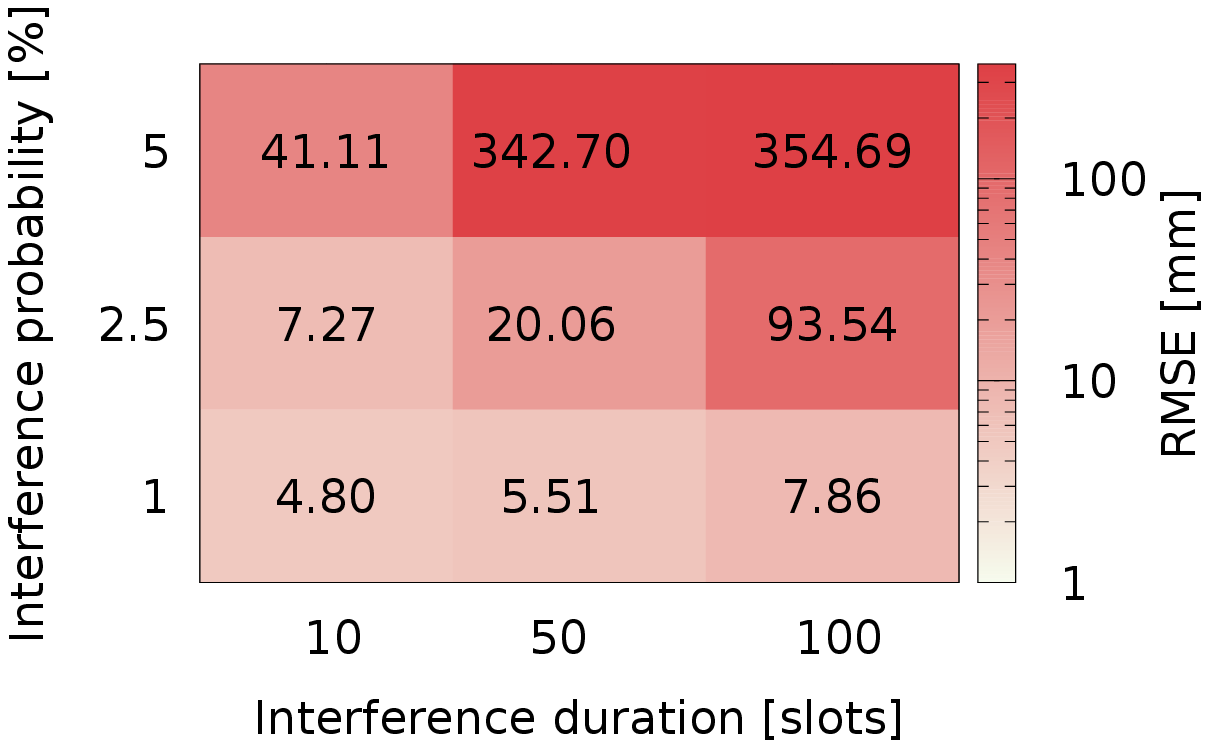}
         \caption{no forecasting - 5 robots}
     \end{subfigure}
     \hfill
     \begin{subfigure}[b]{0.3\textwidth}
         \centering
         \includegraphics[width=\textwidth]{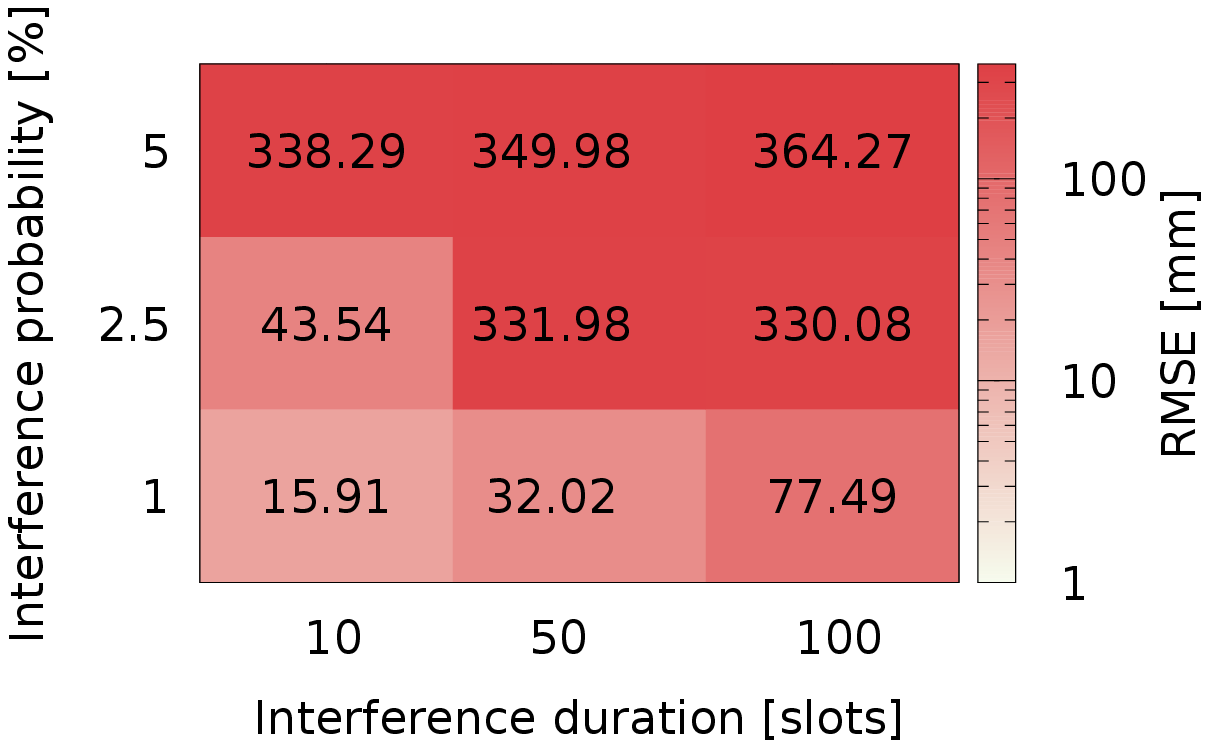}
         \caption{no forecasting - 15 robots}
     \end{subfigure}
     \hfill
          \begin{subfigure}[b]{0.3\textwidth}
         \centering
         \includegraphics[width=\textwidth]{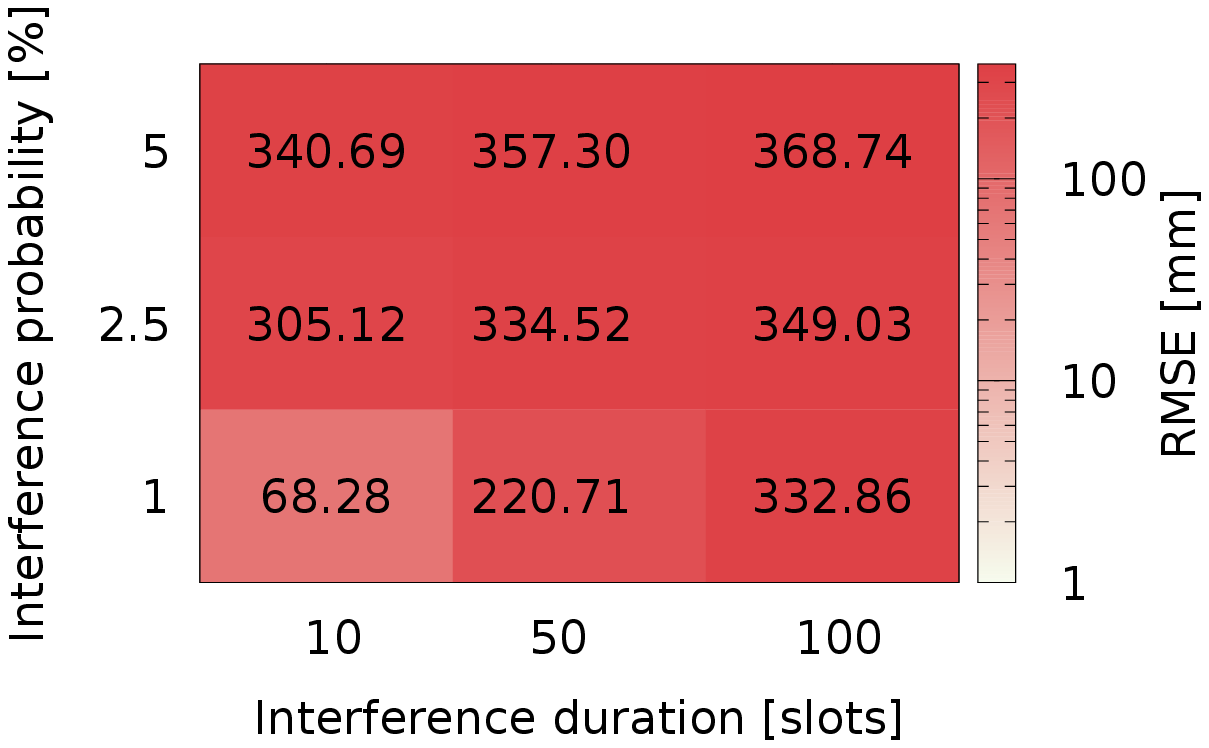}
         \caption{no forecasting - 25 robots}
     \end{subfigure}
     \\
     \begin{subfigure}[b]{0.3\textwidth}
         \centering
         \includegraphics[width=\textwidth]{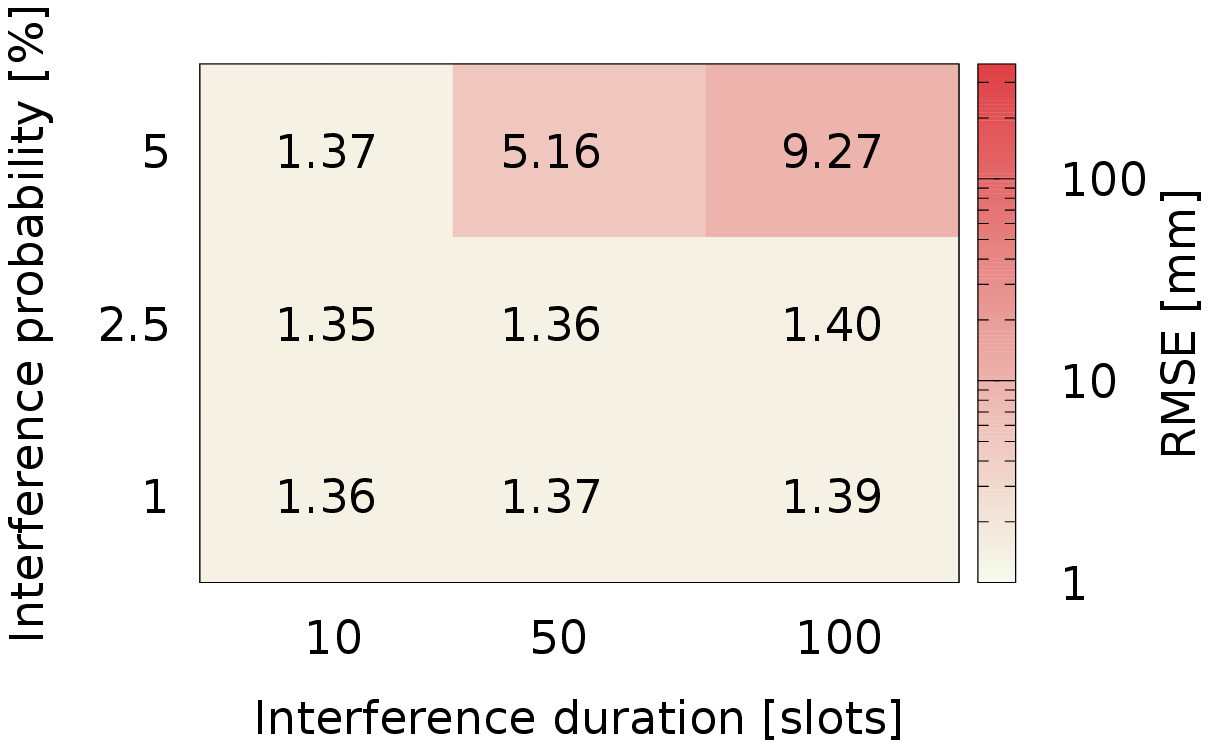}
         \caption{FoReCo - 5 robots}
     \end{subfigure}
     \hfill
     \begin{subfigure}[b]{0.3\textwidth}
         \centering
         \includegraphics[width=\textwidth]{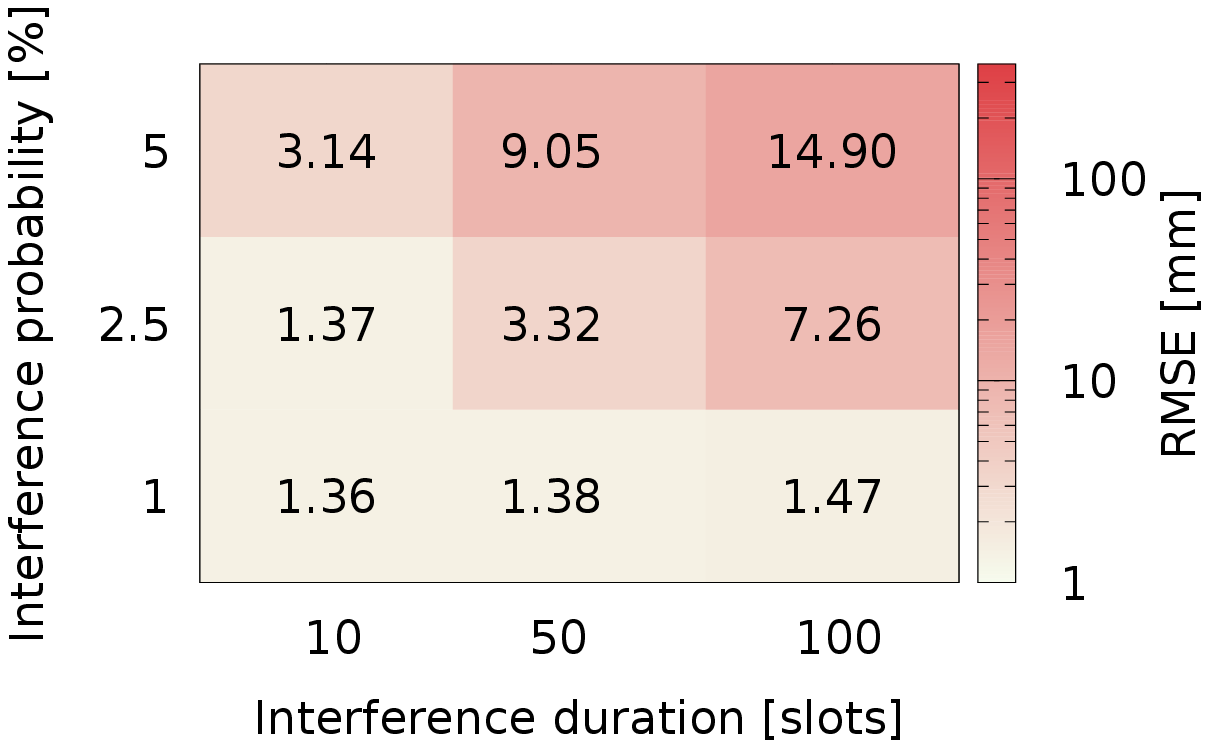}
         \caption{FoReCo - 15 robots}
     \end{subfigure}
     \hfill
          \begin{subfigure}[b]{0.3\textwidth}
         \centering
         \includegraphics[width=\textwidth]{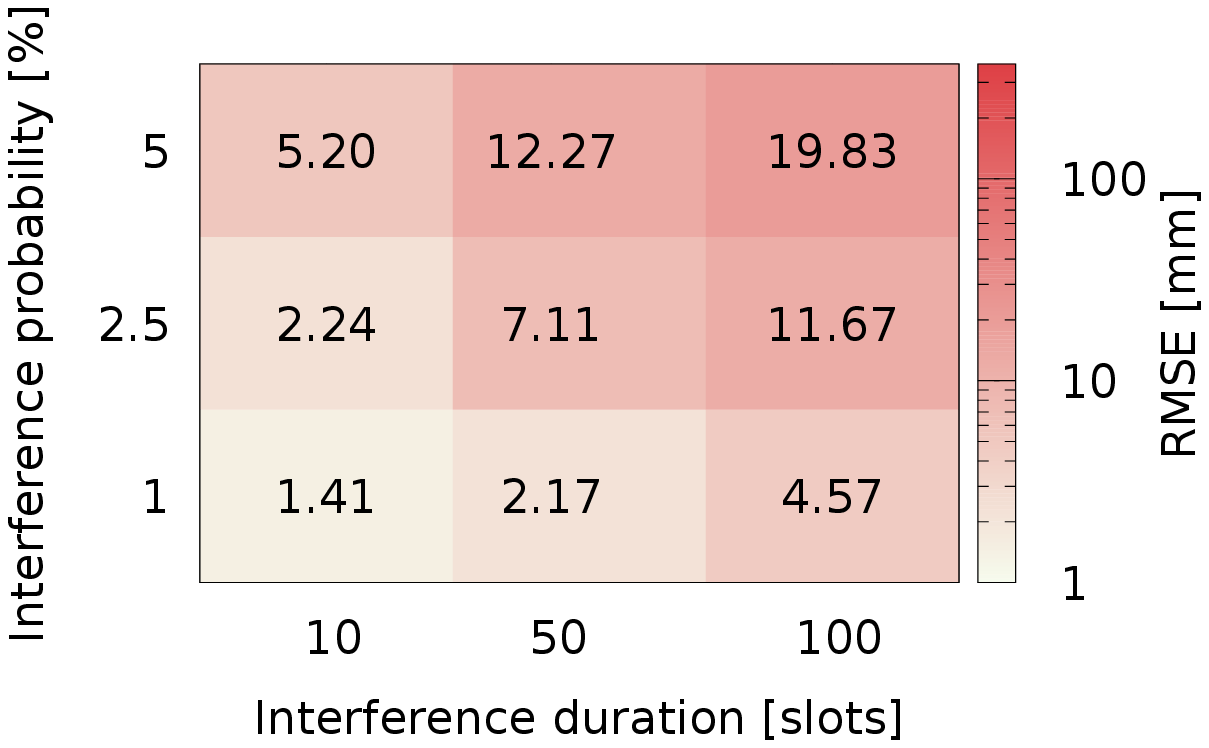}
         \caption{FoReCo - 25 robots}
     \end{subfigure}

\caption[.]{Robot trajectory error upon interference without forecasting (top), and with FoReCo (bottom).}
\label{fig:simulation-error}
\end{figure*}

\subsection{Forecasting accuracy}
\label{subsec:accuracy}
We now evaluate which of the selected forecasting algorithms achieves the highest forecasting
accuracy in the collected datasets.
The VAR algorithm was implemented using
statsmodel v0.12.1, and seq2seq using
Tensorflow 2.1.0.
Fig.~\ref{fig:impact-ahead} shows the RMSE accuracy of each
algorithm as we increase the forecasting window, i.e., how
many consecutive commands are forecasted (a command is
sent each \mbox{$\Omega=20$ ms}). For every algorithm,
we considered a record of the last
\mbox{$R=1,\ldots,20$ commands}, and
Fig.~\ref{fig:impact-ahead} plots the best-performing $R$ parameter
for each algorithm. For the training stage
\mbox{(see~\S\ref{subsec:training})}, we used the
$\alpha=80\%$ of the experienced human operator
data for training, and a $\beta=20\%$ of the
inexperienced operator data for testing.
For seq2seq we resort to the standard hyper-parameter selection:
\mbox{$\eta=0.001$},
\mbox{$\beta_1=0.9$},
\mbox{$\beta_2=0.999$},
\mbox{$\varepsilon=1e-07$}.

Results show that VAR has slightly better accuracy than MA, while seq2seq has the worst performance. It
was expected that VAR would outperform MA since it is designed
for correlated time-series -- like the 6-axis
time-series of the
Niryo One robotic arm. However, seq2seq yielded worse accuracy
than MA due to the vast number of weights
$|\Vec{w}|=163803$ to learn, thus, it did not converge to
an optimal solution. Given the results in 
Fig.~\ref{fig:impact-ahead}, we use the trained
VAR solution as forecasting
technique in the simulations and experiments
in~\S\ref{subsec:simulation} and \S\ref{subsec:experimental},
respectively.

\subsection{Simulation evaluation}
\label{subsec:simulation}
In the following, we evaluate how FoReCo behaves under a
simulated environment with wireless interference.
We consider a transport network with negligible transport
delay, i.e., $D\simeq0$ ms in \textit{Assumption}~\ref{assum},
thus, commands' delays are dominated by the wireless 
delay $\Delta(c_i)\simeq\Delta_W(c_i)$. To derive
$\Delta_W(c_i)$, we resort
to an analytical model of IEEE 802.11 with non-IEEE
interfering sources~\cite{80211analytical}, and
use the parameters reported in~\cite[Table 2]{80211analytical}.
The goal of the simulation validation is two-folded:
($i$) evaluate the precision of the forecasted commands by FoReCo, 
and ($ii$) assess the scalability with up to 25 robotic
arms
sharing a wireless medium with interferences.

Each simulation issues the commands of an inexperienced
human operator and introduces command delays $\Delta_W(c_i)$
following~\cite{80211analytical}.
Fig.~\ref{fig:simulation-error} compares, in the upper and
lower rows, the error experienced by the robot
trajectory when the state-of-the-art solution is used (i.e., repeat
the prior command $\hat{c}_{i+1}=\hat{c}_i$ upon delays),
and when FoReCo recovers the packets as specified
in~\eqref{eq:var}. Since the introduced wireless delay 
$\Delta_W(c_i)$ is a random variable, we repeat each simulation
40 times. Note that, in each simulation, we vary the time and
probability of the active interference.
Each square in the Fig.~\ref{fig:simulation-error}
heatmap illustrates the averaged RMSE of the 40 simulations
done for every pair of interference duration, and probability.
The RMSE is computed over the entire robot trajectory induced by
the inexperienced human operator, and it
considers commands
arriving on time \mbox{$\Delta(c_i)\leq\tau$} and out of time \mbox{$\Delta{c_i}>\tau$},
without using control command forecasting, and with FoReCo
(upper and lower row in~Fig.~\ref{fig:simulation-error},
respectively).

The RMSE error in Fig.~\ref{fig:simulation-error} is
represented in logarithmic scale, and we can appreciate
that FoReCo command recovery constrained the robot
trajectory error below \mbox{9.27, 14.90 and 19.83 mm} for 5,15 and 25 robots on the factory floor, respectively. On the other hand, the no forecasting solution
resulted in an RMSE in the order of $\sim 350$ mm
in the worst cases, no matter the number of robots.
Thus, simulations prove that
($i$)~FoReCo will not exceed errors of 20 mm, and
($ii$)~FoReCo reduces the experienced error by more than one order of magnitude.
In particular, FoReCo reduces by more
than x18 times ($368.74~\text{mm}/19.83~\text{mm}=18.59$) the experienced error
in factory floors of 25 robots
\subsection{Experimental evaluation}
\label{subsec:experimental}
Motivated by the simulation results, we implemented and integrated a prototype of FoReCo within the real remote control system presented in Fig.~\ref{fig:experimental-scenario}. The prototype is following the principles defined in \S\ref{sec:solution},
and uses ROS to interact with the remotely controlled
Niryo One robotic arm.

\subsubsection{Controlled experimental evaluation}
In our first experimental analysis, we manually introduced loss of control commands in order to evaluate the improvements that the prototype offers under a controlled environment. The dataset from the inexperienced user was used to develop a remote controller that was randomly dropping consecutive control commands. Every time consecutive control commands were lost, FoReCo injected predictions from the VAR model. We executed 3 different sets of experiments, where the remote controller randomly introduced 5, 10 or 25 consecutive losses. Each experiment run was 30 seconds, with the robot joint states being recorded in the robot itself.  
  \begin{figure*}[t]
     \centering
     \begin{subfigure}[b]{0.32\textwidth}
         \centering
         \includegraphics[width=\textwidth]{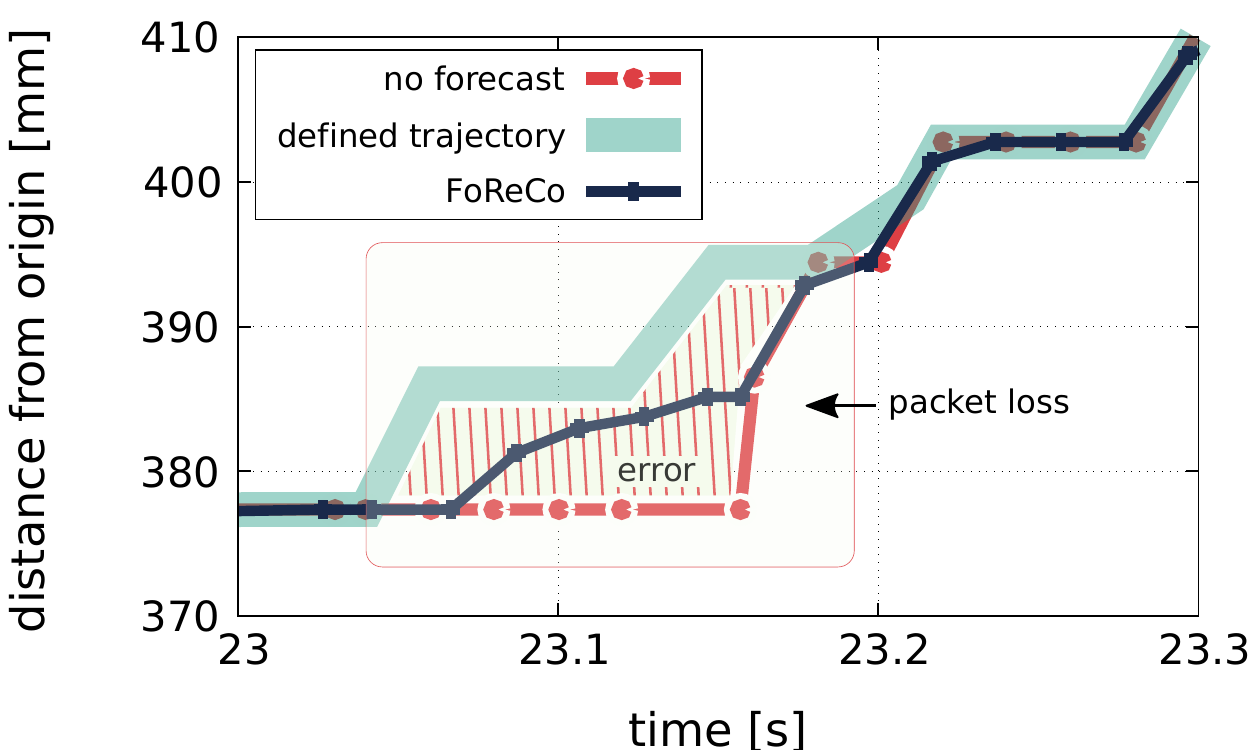}
         \caption{5 consecutive losses}
     \end{subfigure}
     \hfill
     \begin{subfigure}[b]{0.32\textwidth}
         \centering
         \includegraphics[width=\textwidth]{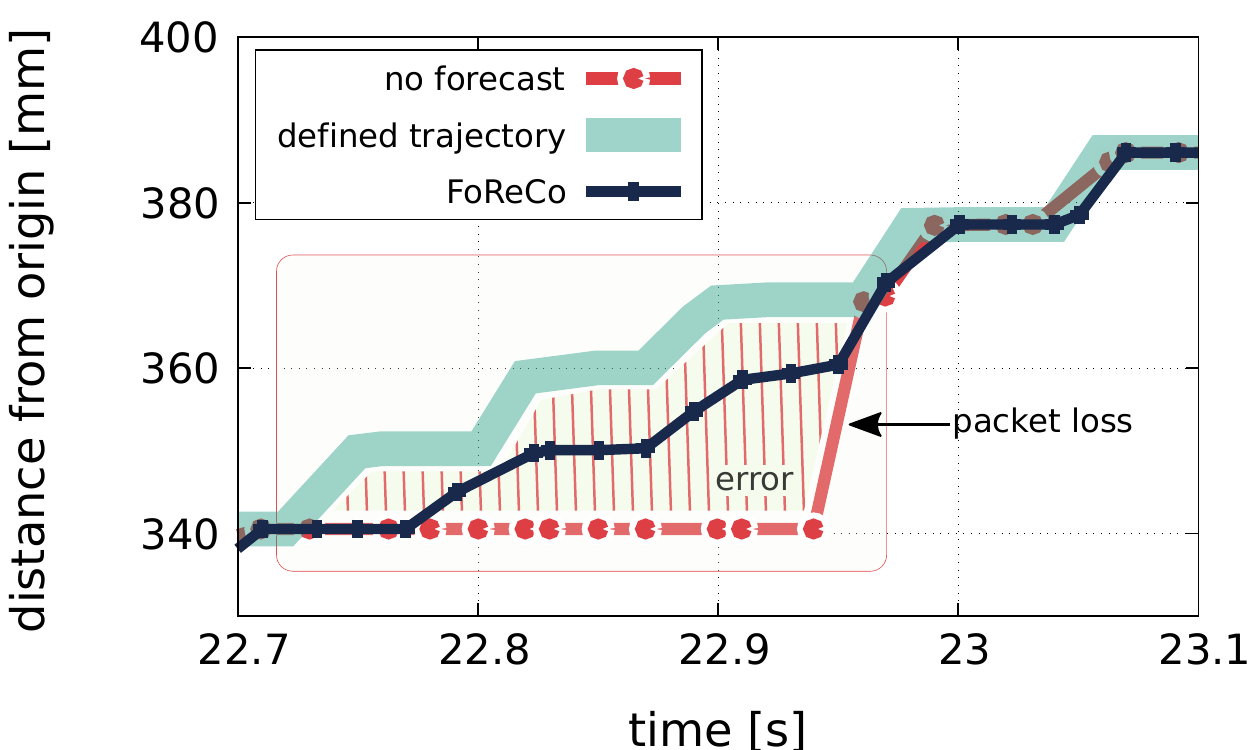}
         \caption{10 consecutive losses}
     \end{subfigure}
     \hfill
     \begin{subfigure}[b]{0.32\textwidth}
         \centering
         \includegraphics[width=\textwidth]{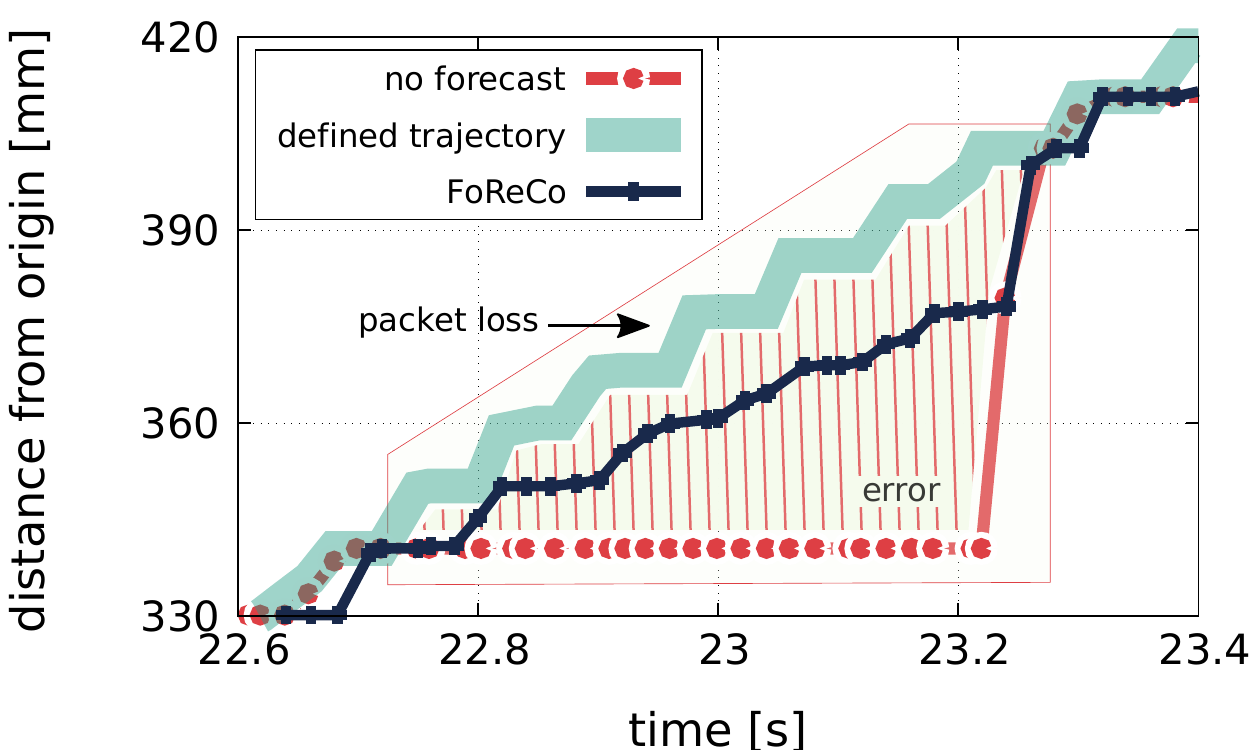}
          \caption{25 consecutive losses}
     \end{subfigure}
\caption[.]{Robot trajectory with controlled packet losses using
no forecasts, and FoReCo.}
\label{fig:experimental_results}
\end{figure*}

Fig.~\ref{fig:experimental_results} shows the trajectories followed by the robot arm when consecutive control commands were lost for the case of no forecasts, and the FoReCo solution.
The results show that FoReCo can mitigate the negative effects of lost commands by minimizing the trajectory error in the 3 different sets of experiments. Moreover, results show that the RMSE for all three experimental sets is between \mbox{1.35 mm}
and \mbox{9.27 mm}, which makes it consistent with the simulation results obtained for 5 stations (see Fig.~\ref{fig:simulation-error}(d)). However, Fig.~\ref{fig:experimental_results}(c) shows
how FoReCo deviates more and more from the
defined trajectory as the number of consecutive
losses increase (in between second 23 and 23.2),
since VAR builds its forecasts $\hat{c}_{i+1}$
using prior forecasted commands
\mbox{$\hat{c}_i, \hat{c}_{i-1}, \ldots, \hat{c}_{i-R}$}.
Thus, the prediction error propagates --
see~\eqref{eq:var}.

\subsubsection{Jammed experimental evaluation}
\begin{figure}[t]
    \centering
    \includegraphics[width=\columnwidth]{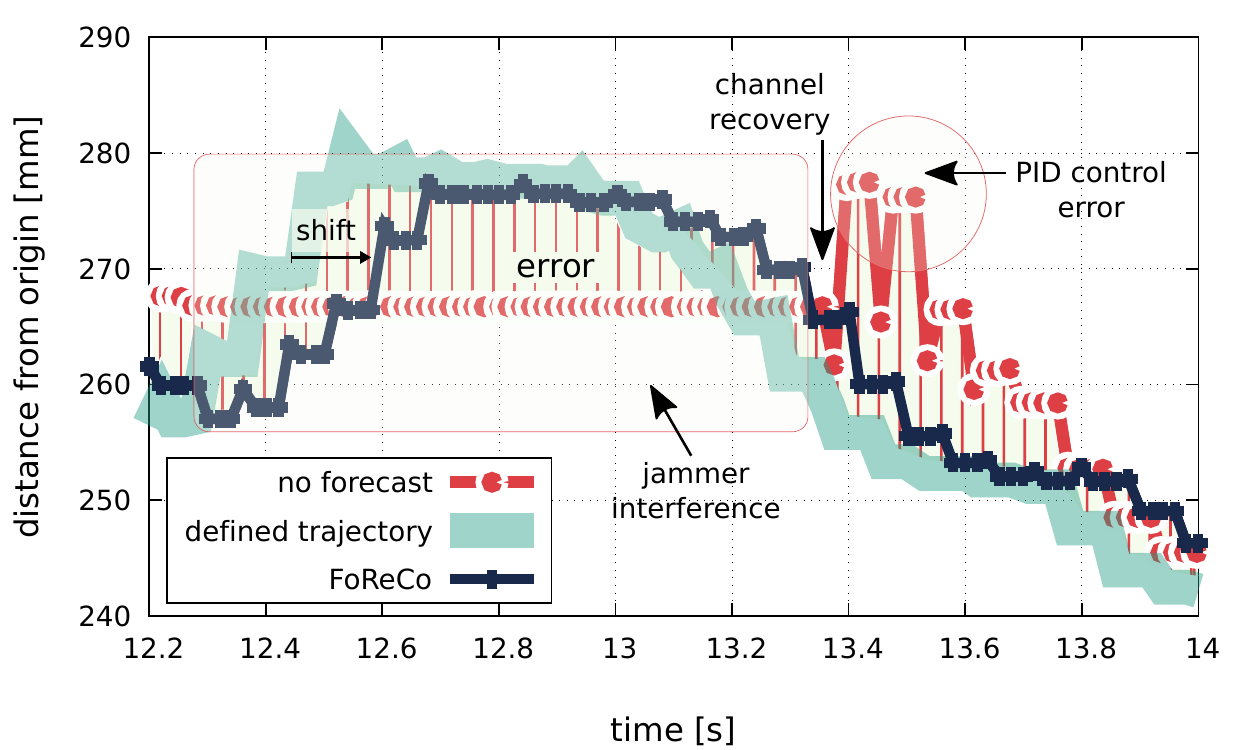}
    \caption{Robot trajectory upon IEEE 802.11 jammer interference
    }
    \label{fig:jammer-trajectories}
    \vspace{-1em}
\end{figure}

In our second experimental analysis, we emulate a realistic interference scenario where the Silvercrest Wireless device is used to transmit synchronized radio waves in the same frequency as the robot, introducing unpredictable network delays $\Delta(c_i)$ and
packet losses -- same as the analytical model~\cite{80211analytical}
used in prior simulations in \S\ref{subsec:simulation}.
The dataset from the inexperienced user was used in order to develop a remote controller that sent the control commands over the jammed wireless network. Every time control command(s) were lost or delayed due to the interference $\Delta(c_i)>\tau$, FoReCo injected predictions from the VAR model. The experiment was executed for 30 seconds and in the robot, we recorded the robot joint states. 

Fig.~\ref{fig:jammer-trajectories} presents the trajectories followed by the robot arm when the wireless channel was interfered by the jammer. The results show how FoReCo reduces by more than x2 times the trajectory error RMSE 
(from \mbox{18.91 mm}
to \mbox{8.72 mm})  -- compared to the bare Nyrio One solution without forecasting.
In the interval between \mbox{12.4 and 12.6 seconds}, FoReCo starts to deviate (shift) from the defined trajectory in an identical way to the controlled experiments described above. In addition, an interesting observation is how the robot trajectory behaves upon channel recovery.
In the latter case, the Niryo One ROS MoveIt PID controller takes around \mbox{400 ms}
to stabilize the trajectory (from second 13.5 to 13.8),
since the PID controller receives repeated commands
\mbox{$c_{i+1}=c_i, \forall i\in[12.4,13.4]$}
during more than a second, resulting in the highlighted error
in Fig.~\ref{fig:jammer-trajectories}.

\subsubsection{Training and inference times}
Since Niryo One is equipped with a Raspberry~Pi3, which is limited in terms of computing power, we executed a set of experiments to measure both training and inference times of VAR whenever executed in the robot itself. While training takes around 5.99$\pm$0.06 min to train using the experienced human operator dataset, the inference (i.e., prediction of the commands) only takes 1.60$\pm$0.16 ms. These results show that the current hardware of the robot is sufficient to not only accommodate FoReCo within the constrains of our prototype but also to support more stringent applications with control loops close to 2 ms. Note that training is only required when we need to build or update the model, which time profiling is presented in Table~\ref{tab:training-profiling}.

\begin{table}[!ht]
\centering
\caption{Time profiling of FoReCo training in Niryo One}
\label{tab:training-profiling}
\begin{tabular}{|p{1.8cm}|p{1.0cm}|p{1.4cm}|p{1.3cm}|p{1.1cm}|}
\hline
              & \textbf{Load Data (s)}   & \textbf{Down Sampling (s)} & \textbf{Check Quality (s)} & \textbf{Training Model (s)} \\ \hline
\textbf{Raspberry Pi3 (Robot)} & 1.95 $\pm$ 0.02 & 0.26 $\pm$ 0.007  & 306.38 $\pm$ 3.15 & 50.98 $\pm$ 0.54 \\ \hline
\end{tabular}
\end{table}

For comparison, Table~\ref{tab:training-and-inference-comparison} also presents training and inference times in different equipment: \textit{(i)} NVIDIA Jetson Nano which can be co-located with Niryo One; \textit{(ii)} a laptop equipped with a 2nd gen Intel~Core~i7 and 6GB~RAM representing the user equipment; and \textit{(iii)} a local server with two Intel(R)~Xeon(R)~CPU~E5-2620~v4@2.10GHz and 64GB~RAM representing the Edge to offload training and inference tasks.

\begin{table}[!ht]
\centering
\caption{Training and inference times in different equipment}
\label{tab:training-and-inference-comparison}
\begin{tabular}{|p{3.6cm}|p{1.8cm}|p{2.1cm}|}
\hline
                   & \textbf{Training (min)}  & \textbf{Inference (ms)}  \\ \hline
\textbf{Raspberry Pi3 (Robot)} & 5.99 $\pm$ 0.06 & 1.60 $\pm$ 0.16 \\ \hline \hline
\textbf{NVIDIA Jetson Nano (Robot)} & 1.31 $\pm$ 0.01 & 0.61 $\pm$ 0.28 \\ \hline
\textbf{Laptop (UE)} & 0.36 $\pm$ 0.01 & 0.22 $\pm$ 0.10 \\ \hline
\textbf{Local Server (Edge)} & 0.23 $\pm$ 0.007 & 0.0001 $\pm$ 0.00003 \\ \hline
\end{tabular}
\end{table}

\section{Discussion}
\label{sec:discussion}
In this section we discuss the results, analyzing how they can be interpreted from the perspective of real-time wireless networked control systems. In addition, possible future directions are identified.

\subsection{Overall performance and applicability}
FoReCo decreased the trajectory error of a robot manipulator by \mbox{up to x18 times} in simulation, and \mbox{up to x2 times} in experimentation, when operating on a   pick-and-place dataset. Results show that FoReCo can achieve high precision and suggest that industrial manipulation applications (e.g., assembly, pick-and-place) can benefit from adopting the proposed method. Results also demonstrate that the prototype of FoReCo can be easily attached to robot manipulators without robot-specific modifications. Furthermore, results show how predictive control can improve the reliability of real-time remote control over IEEE 802.11 wireless network. 

\subsection{Coexistence with emerging wireless technologies}
Existing industrial wireless technologies cannot meet all of the remote control requirements (such as, 99.999\% of reliability, 2-20ms latency, 100-200 Mbps data rate). Although emerging technologies such as WiFi~6E and 5G are expected to be a key enabler in this regard by increasing the levels of reliability in industrial wireless, the uncertain and time varying wireless channel continues to be an issue for wireless technologies. The fact that FoReCo does not make any assumptions about the wireless channel and the network delays makes it applicable also to emerging wireless such as mmWave or 5G. In addition, it is very likely that legacy systems will leverage previous technologies which will operate for years to come, and FoReCo can offer them the needed reliability for running remote control applications.

\subsection{Real-time Path Tracking Predictions}
Concerning the ML algorithm for repetitive reference trajectories, given the performance of VAR, our future work will consider exponential smoothing methods, and Vector Autoregression Moving Average (VARMA). The latter method combines the benefits of both MA and VAR to prevent saw-teeth oscillations, and anticipate faster the increases/decreases of the time-series. It is worth noting that, although VAR performed well on the pick-and-place dataset, a deviation from the defined trajectory is witnessed when the number of consecutive commands losses increases.
Future versions of FoReCo may mitigate
large periods of consecutive losses by
incorporating delayed commands that
did not arrive on time, i.e., commands
$c_i$ with $\Delta(c_i)>\tau$ could
be used instead of the predicted one
$\hat{c}_i$, to infer commands after
\mbox{$n\Omega$ ms}. In other words,
based on \eqref{eq:command} we could
use $\hat{c}_{n}=f(\hat{c}_{n-1},\ldots,c_i,\ldots)$.

\subsection{Predictions at the edge of the network}
In addition to improving the model accuracy and precision, an edge-based version of FoReCo can be considered where the VAR algorithm will always base its predictions on the real control commands $\hat{c}_i=f(\{c_j\}_{i-R}^{i-1})$. However, this approach is a bit more disruptive because an edge-based version of FoReCo indicates that the predictions will need to traverse the interfered wireless channel. Hence, FoReCo will need to piggyback the predictions together with real-time control commands.
Piggybacking predictions requires modifications of the robot drivers to use them whenever a control command does not arrive on time. We leave such an option for future work. 

\subsection{An extra resilience layer}
As stated in \S\ref{sec:related-work}, FoReCo targets a solution from the network perspective, in opposition to already existing related work. However, the two approaches are not mutually exclusive and, consequently, they can be used together in order to provide even higher levels of reliability and precision when performing remote control of robotic manipulators. In doing so, in a full-fledged solution, several recovery mechanisms can be envisioned as a resilience stack where all layers work together in a fail-over fashion.

\subsection{Extension to other robotic systems}
Although FoReCo explicitly targeted robotic manipulators, the solution described in this work can be easily extended to many other robotic systems that implement open- or close-loop between themselves and the remote controlling system. For example, remotely controlled AGVs within the manufacturing plant. Similarly to the robotic manipulators, missing commands are predicted and injected into the AGV so that it can smoothly follow its trajectory. Still, FoReCo requires that performed tasks are somehow periodic so that the corresponding dataset can be created, and training and inference tasks executed with acceptable levels of precision. 

\section{Conclusions}
\label{sec:conclusions}
This paper presents FoReCo, a forecast-based recovery mechanism for real-time robot remote control. The FoReCo prototype uses VAR and ROS,
and its
performance has been assessed in a commercial research robotic arm remotely controlled over IEEE 802.11 wireless channels under the presence of interference.
We also validate FoReCo through simulation and assess its performance in a real testbed.
Results show that FoReCo provides high precision by achieving a trajectory  error 
below \mbox{19.83 mm} in simulation, and of \mbox{8.72 mm} in experimentation with a commercial research robotic arm.
This means that FoReCo reduces
by up to x18 and x2 times the
trajectory error in simulation and
experimentation.
Moreover, FoReCo is lightweight and can be deployed in the hardware already available in several solutions.

As follow-up work, we plan to adapt solutions
of wireless controlled AGVs~\cite{avg-1}\cite{avg-2}
to remotely controlled robotic arms,
and compare their performance
against FoReCo in \mbox{IEEE 802.11} wireless channels.



\section*{Acknowledgment}
This work has been partially funded by European Union’s Horizon 2020 research and innovation programme under grant agreement No 101015956, and the Spanish Ministry of Economic Affairs and Digital Transformation and the European Union-NextGenerationEU through the UNICO 5G I+D 6G-EDGEDT and 6G-DATADRIVEN.

\appendix
\label{appendix}

In the following, we present some theoretical results
about the expected delay of a command, and the causality
assumption in the IEEE~802.11 scenario considered in this
paper.

\begin{lemma}
    A control command $c_i$ traversing a transport network,
    and an IEEE~802.11 wireless link under interference
    will experience an average delay satisfying
    \begin{equation}
        \mathbb{E}\left[ \Delta(c_i) \right] \leq
            D + \frac{1}{1-a_{m+2}}\sum_{j=0}^{m+1} a_j\cdot \mathbb{E}_j\left[ \Delta_W(c_i) \right],\quad \forall c_i
            \label{eq:corollary}
    \end{equation}
    with probability $1-a_{m+2}$, and
    \begin{equation}
        \mathbb{E}\left[ \Delta(c_i) \right]=\infty, \quad \forall c_i
        \label{eq:corollary-2}
    \end{equation} with probability $a_{m+2}$.
    $m+2$ being the maximum number of allowed re-transmissions in
    IEEE~802.11 wireless links.
    \label{lemma:upper-bound}
\end{lemma}
\begin{proof}
    In case a command is not lost in the IEEE~802.11
    wireless link (less than $m+2$ re-transmissions),
    if we take the law of total probability
    and the analytical model in~\cite{80211analytical},
    the average wireless delay is
    \begin{equation}
        \mathbb{E}\left[ \Delta_W(c_i) \right] = \sum_{j=0}^{m+1} a_j\cdot \mathbb{E}_j\left[ \Delta_W(c_i) \right], \quad \forall c_i
        \label{eq:proof-corollary}
    \end{equation}
    Note that this happens with probability $1-a_{m+2}$.
    Therefore, if we foresee that a command is not lost,
    we have to rescale~(\ref{eq:proof-corollary}) by
    $\tfrac{1}{1-a_{m+2}}$, i.e., the probability that
    a command is not lost.
    
    Since we know that $\Delta(c_i)=\Delta_T(c_i)+\Delta_W(c_i)$,
    if we take the expectation at both sides of the equatlity,
    use Assumption~\ref{assum}, and the rescaled version
    of~(\ref{eq:proof-corollary}); we obtain~(\ref{eq:corollary}).
    
    According to~\cite{80211analytical}, (\ref{eq:corollary-2}) 
    holds because a packet is lost $\Delta_W(c_i)=\infty$
    in a IEEE~802.11 wireless link with probability $a_{m+2}$.
\end{proof}

\begin{corollary}
    A control command $c_i$ traversing a transport network,
    and an IEEE~802.11 wireless link under interference,
    experiences an unbounded delay, that is
    \begin{equation}
        \mathbb{P}\Big( \Delta(c_i)>K,\ \forall K\in\mathbb{R} \Big) > 0
    \end{equation}
    \label{corol:unbound}
\end{corollary}
\begin{proof}
    In particular, Lemma~\ref{lemma:upper-bound} says that
    \mbox{$\mathbb{P}\Big( \Delta(c_i)>K,\ \forall K\in\mathbb{R} \Big) =a_{m+2}$}.
\end{proof}

\begin{lemma}
    In IEEE~802.11 wireless links, the causality assumption
    \begin{equation}
        |\Delta(c_{i+1}) - \Delta(c_i)| \leq |g(c_{i+1})-g(c_i)|, \quad \forall c_{i+1}, c_i
        \label{eq:causality}
    \end{equation}
    only holds on average with probability 
    $\sum_{j=0}^{m+1}a_j^2$.
    \label{lemma:causality-only}
\end{lemma}
\begin{proof}
    We prove the lemma by cases, namely considering the
    different combinations of required re-transmissions
    of commands $c_i$ and $c_{i+1}$.
    
    If either command $c_i$ or $c_{i+1}$ is lost
    ($m+2$ re-transmissions),
    then we have with probability $a_{m+2}$ that
    equation~\ref{eq:causality} does not hold, since
    either $\Delta_W(c_i)=\infty$ or
    $\Delta_W(c_{i+1})=\infty$;
    
    If $c_{i+1}$ has $j_2$ RTX,
    and $c_i$ has $j_1$ RTX (with $j_1<j_2$), then
    \begin{equation}
        \Delta_W(c_{i+1})-\Delta_W(c_i) = |\Delta_W(c_{i+1}-\Delta_W(c_i)|=\leq g(c_{i+1})-g(c_i)
    \end{equation}
    We can take the expectation on the left side hand
    and obtain:
    \begin{multline}
        \mathbb{E}\left[ \Delta_W(c_{i+1}) \right] - \mathbb{E}\left[ \Delta_W(c_i) \right] =\\
        T_s+j_2T_{col}+\tilde{\sigma}\sum_{k=0}^{j_2}\frac{W_k-1}{2} - T_s-j_1T_{col}-\tilde{\sigma}\sum_{k=0}^{j_1}\frac{W_k-1}{2}=\\
        (j_2-j_1)T_{col}+\tilde{\sigma}\sum_{k=j_1+1}^{j_2}\frac{W_k-1}{2}
        \label{eq:diff-rtx}
    \end{multline}
    with $T_s$ the transmission time,
    $T_{col}$ the collision time,
    $\tilde{\sigma}$ the average slot time,
    and $W_k$ the $k$\textsuperscript{th} back-off window.
    Based on (\ref{eq:diff-rtx}), the causality
    assumption does not hold, since
    \begin{multline}
        \exists~T_s,T_{col},\tilde{\sigma},W_k:\\
        (j_2-j_1)T_{col}+\tilde{\sigma}\sum_{k=j_1+1}^{j_2}\frac{W_k-1}{2} > g(c_{i+1}) - g(c_i)
    \end{multline}
    The same reasoning applies in the case $j_1>j_2$.
    
    If command $c_i$ required same re-transmissions as
    command $c_{i+1}$ (i.e. $j_1=j_2$), then (\ref{eq:diff-rtx})
    $\mathbb{E}\left[ \Delta_W(c_{i+1}) \right]=0$,
    and the causality assumption~(\ref{eq:causality}) holds.
    This event occurs with probability $\sum_{j=0}^{m+1}a_j^2$.
\end{proof}

\begin{corollary}
    In IEEE~802.11 wireless links, the causality
    assumption
    \begin{equation}
        |\Delta(c_{i+1}) - \Delta(c_i)| \leq |g(c_{i+1})-g(c_i)|, \quad \forall c_{i+1}, c_i
        \label{eq:causality}
    \end{equation}
    does not hold.
    \label{corol:no-causality}
\end{corollary}
\begin{proof}
    The causality assumption does not hold with probability 1 
    (see Lemma~\label{Lemma:causality-only}), therefore,
    the causality assumption does not hold in IEEE~802.11
    wireless links.
\end{proof}

\bibliographystyle{IEEEtran}
\bibliography{bibliography.bib}


\begin{IEEEbiography}[{\includegraphics[width=1in,height=1.25in,clip,keepaspectratio]{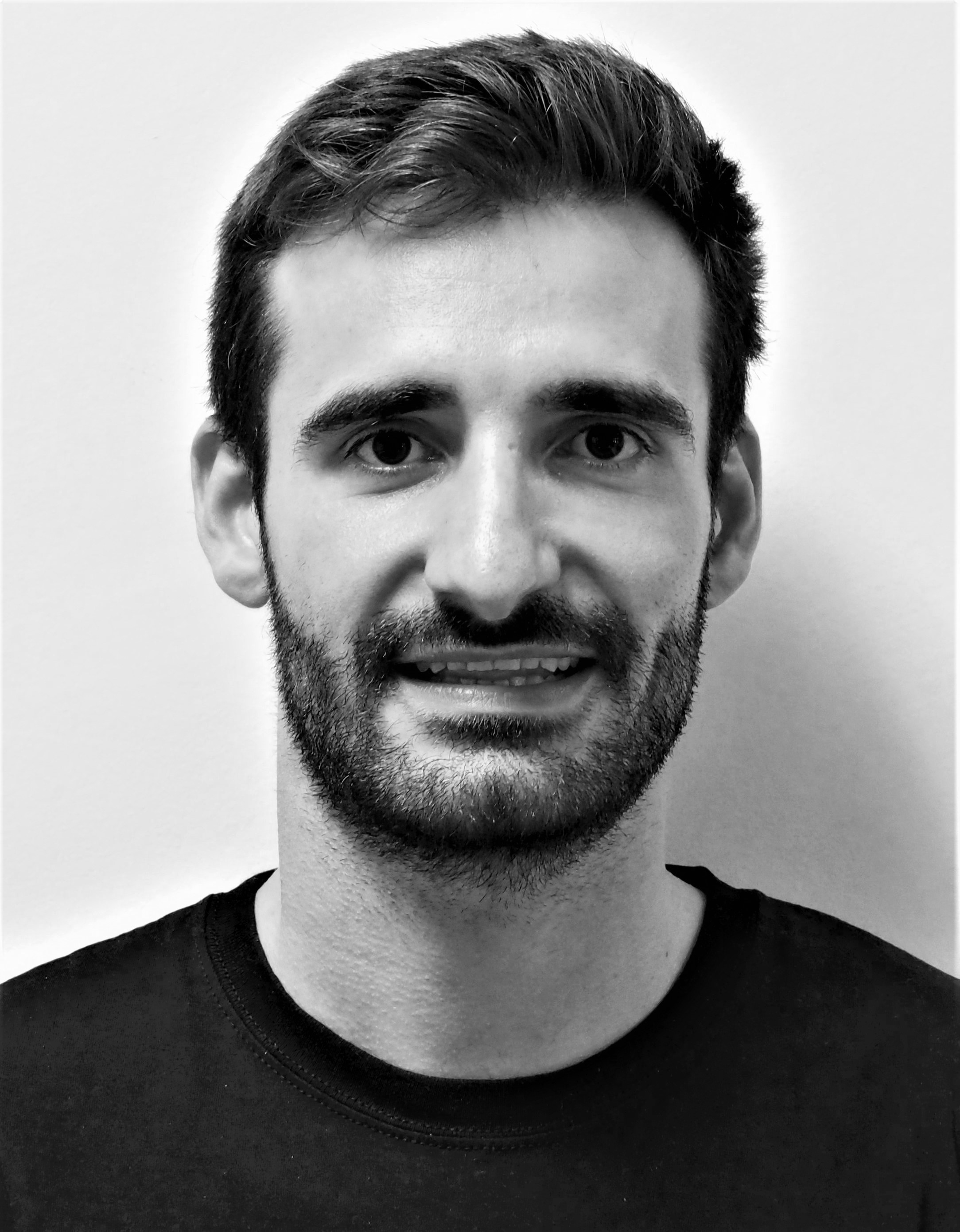}}]{Milan Groshev}
received the B.S. degree in telecommunication engineering from the Saints Cyril and Methodius University of Skopje, Macedonia in 2008 and the M.S. degree in telecommunication engineering from the Politecnico di Torino, Turin, Italy in 2016. He is currently pursuing the Ph.D. degree in telematics engineering at University Carlos III Madrid (UC3M), Spain.

His doctoral research investigates the integration of Edge and Fog in virtual environments with objective to build lightweight, low cost and smarter robots.
\end{IEEEbiography}

\begin{IEEEbiography}[{\includegraphics[width=1in,height=1.25in,clip,keepaspectratio]{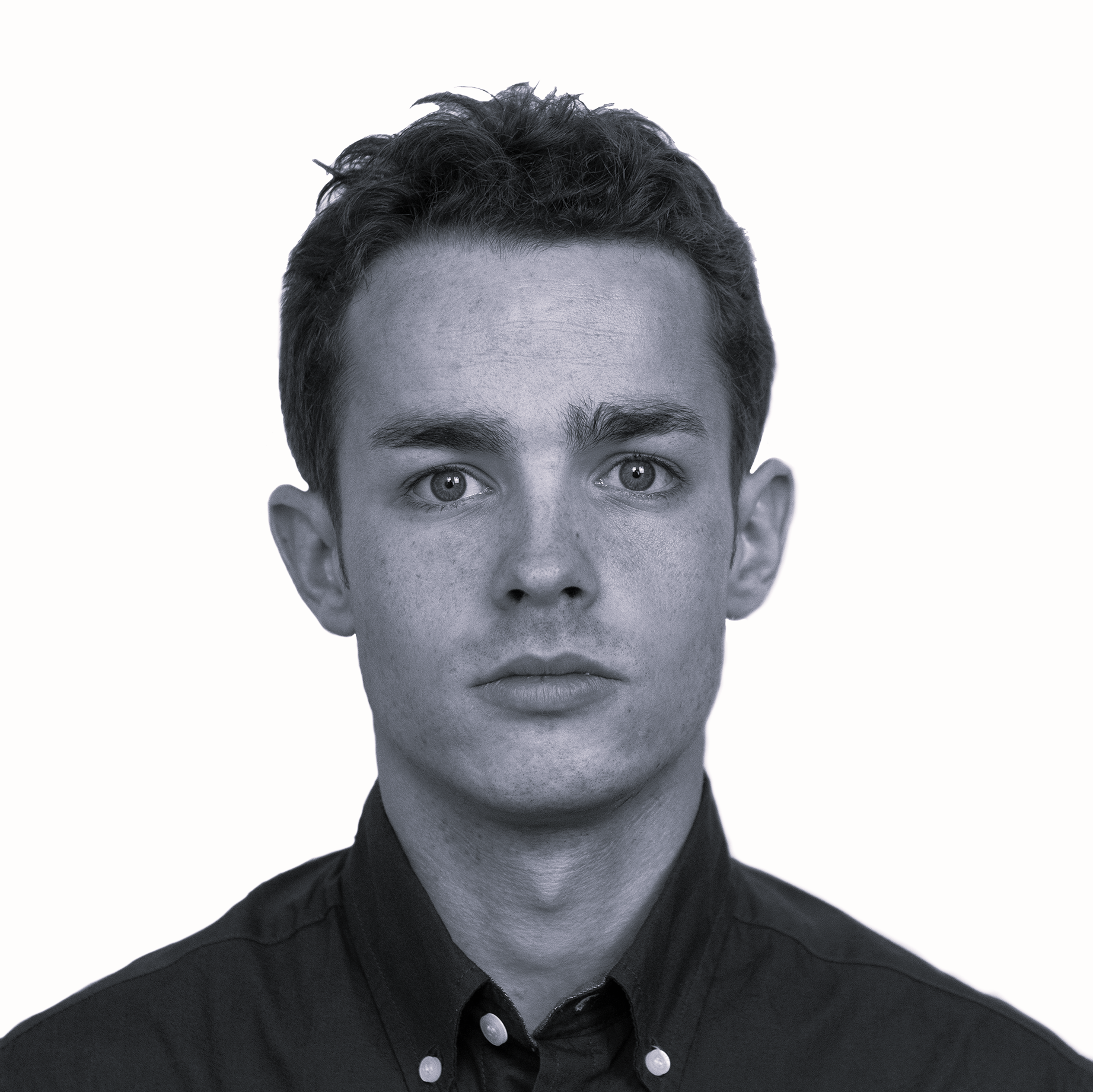}}]{Jorge Mart\'in P\'erez}
obtained a B.Sc in mathematics, and
a B.Sc in computer science, both at Universidad
Autónoma de Madrid (UAM) in 2016. He obtained his M.Sc. and Ph.D in
Telematics from Universidad Carlos III de Madrid (UC3M) in 2017 and 2021, respectively.
His research focuses in optimal resource allocation in
networks, and since 2016 he participates in
EU funded research projects in UC3M Telematics department.
\end{IEEEbiography}

\begin{IEEEbiography}
[{\includegraphics[width=1in,height=1.25in,clip,keepaspectratio]{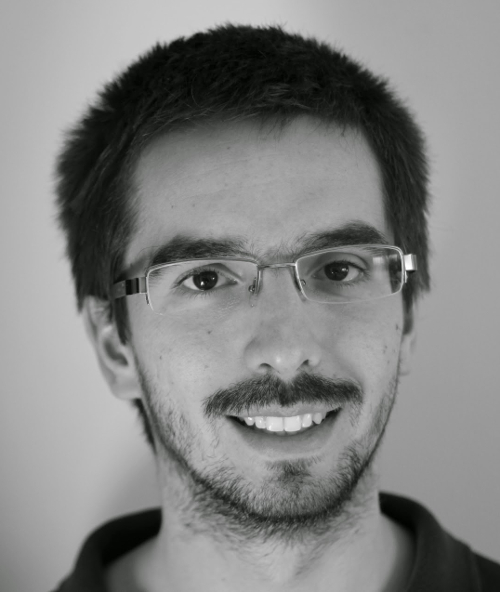}}]
{Carlos Guimar\~{a}es}
is currently a Senior Technologist at ADLINK Technology (France) where he is developing data-centric networking solutions. Prior to that, he had worked as a Postdoctoral Researcher at Universidad Carlos III de Madrid (Spain), having received the M.Sc. degree in computer and telematics engineering from the Universidade de Aveiro (Portugal) in 2011, and the Ph.D. degree in computer science, in 2019, under the scope of MAP-i Doctoral Program (Portugal).
\end{IEEEbiography}

\begin{IEEEbiography}
[{\includegraphics[width=1in,height=1.25in,clip,keepaspectratio]{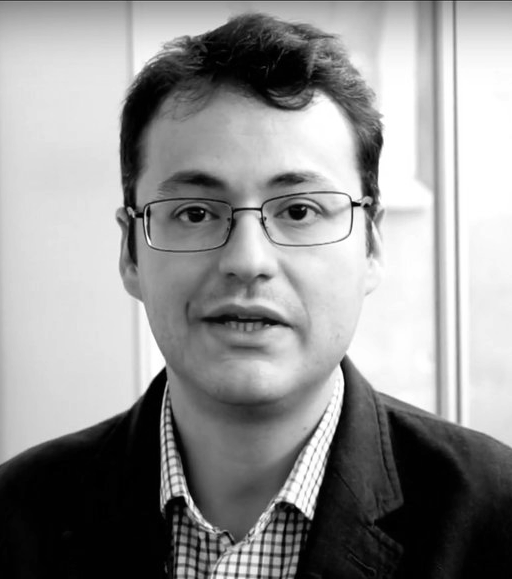}}]
{Antonio de la Oliva}
received his telecommunications engineering degree in 2004 and his Ph.D. in 2008 from the Universidad Carlos III Madrid (UC3M), Spain, where he has been an associate professor since then. 

He is an active contributor to IEEE 802 where he has served as Vice-Chair of IEEE 802.21b and Technical Editor of IEEE 802.21d. He has also served as a Guest Editor of IEEE Communications Magazine. He has published more than 30 papers on different networking areas.
\end{IEEEbiography}

\begin{IEEEbiography}
[{\includegraphics[width=1in,height=1.25in,clip,keepaspectratio]{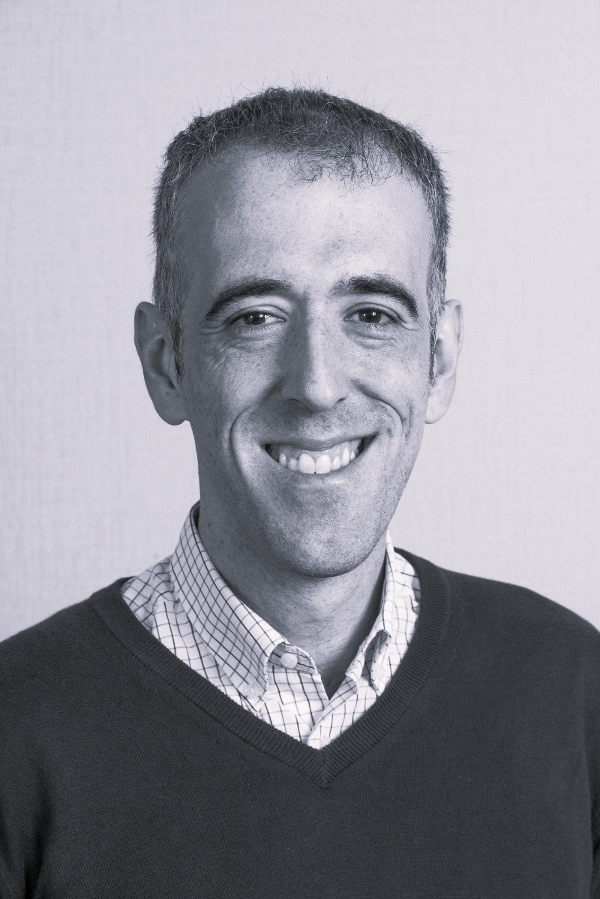}}]
{Carlos J. Bernardos}
received a Telecommunication Engineering degree in 2003, and a PhD in Telematics in 2006, both from the University Carlos III of Madrid, where he worked as a research and teaching assistant from 2003 to 2008 and, since then, has worked as an Associate Professor. His research interests include IP mobility management, network virtualization, cloud computing, vehicular communications and experimental evaluation of mobile wireless networks. He has published over 70 scientific papers in international journals and conferences. He has participated in several EU funded projects, being the project coordinator of 5G-TRANSFORMER and 5Growth.
\end{IEEEbiography}

\end{document}